\documentclass[useAMS,usenatbib]{mn2e}

\usepackage{amsmath}
\usepackage{amssymb}
\usepackage{graphicx}

\title[Maximum likelihood estimation of stellar magnetic fields]{Analytical maximum
likelihood estimation of stellar magnetic fields}
\author[M. J. Mart\' inez Gonz\'alez]{M. J. Mart\' inez Gonz\'alez$^{1,2}$, R.
Manso Sainz$^{1,2}$, A. Asensio Ramos$^{1,2}$, and L. Belluzzi$^{1,2}$\\
$^{1}$Instituto de Astrof\' isica de Canarias, V\' ia L\'actea s/n, 38205, La
Laguna, Tenerife, Spain\\
$^{2}$ Departamento de Astrof\'isica, Universidad de La Laguna, 38205, La
Laguna, Tenerife}
\begin{document}

\date{}

\pagerange{\pageref{firstpage}--\pageref{lastpage}} \pubyear{2002}

\maketitle

\label{firstpage}

\begin{abstract}
The polarised spectrum of stellar radiation encodes valuable information on the
conditions of stellar atmospheres and the magnetic fields that permeate them. In
this paper, we give explicit expressions to estimate the magnetic field
vector and its associated error from the observed Stokes parameters. We study
the solar case 
where specific intensities are observed and then the stellar case, where we 
receive the polarised flux. In this second case, we concentrate on the explicit
expression 
for the case of a slow rotator with a dipolar magnetic field geometry. Moreover,
we also 
give explicit formulae to retrieve the magnetic field vector from the LSD
profiles without 
assuming mean values for the LSD artificial spectral line.
The formulae have been obtained assuming that the spectral lines
can be described in the weak field regime and using a maximum likelihood
approach. The errors are recovered by means of the hermitian matrix. The
bias of the estimators are analysed in depth.
\end{abstract}
  

\section{Introduction}
The most precise measurements of stellar magnetic fields are based on the
observation and the interpretation of polarisation in spectral lines. Most of
the magnetic fields detected
in stars at different phases of evolution have been found by means of the Zeeman
effect, which generates linear and circular polarisation in the presence of a
magnetic field
\citep[e.g.,][]{wade00,bagnulo02,jordan05,aznar_cuadrado04,otoole05,silvester09,
leone11}.
Here we present analytical expressions for inferring the magnetic field vector
from the observed Stokes
profiles induced by the Zeeman effect in the weak field approximation.

The weak field approximation is broadly applied for the inference of solar and
stellar magnetic fields from the observation
of the Stokes profiles. It is an analytical solution to the radiative transfer
equation whose main basic assumption is that the magnetic field is sufficiently
weak throughout the whole region of the atmosphere where a spectral line is
formed. Although simple, this approximation is
very useful since non-magnetic mechanisms usually dominate the shape of
spectral lines, in both the solar and the stellar case. 

For example, in the quiet Sun, which occupies the vast majority of the solar
surface (far from the sunspots that harbour very strong fields), the spectral
lines are well described by the weak field approximation. This has allowed the
success of many synoptic magnetographs like those
of Big Bear \citep{spirock01,varsik95}. It is even used to produce modern vector
magnetograms like those obtained with the IMaX instrument 
\citep{imax11} onboard the Sunrise balloon \citep{sunrise10}. The weak field
approximation 
has also been used to diagnose the
chromosphere \citep{merenda06,asensio_trujillo_hazel08} due to the 
enhanced thermal width of the spectral lines. 

In night-time spectro-polarimetry, the weak field approximation is at the
base of the least-squares deconvolution \citep[LSD; ][]{donati97}, the most 
successful technique used to detect and measure magnetic fields in solar-type
stars or
other stars in which the polarisation signal per spectral line is well below the
noise level. Other works
have used this approximation to diagnose magnetic fields in a large variety of
stellar objects. A limited selection includes some recent works on central stars
of planetary nebulae \citep{jordan05, leone11}, white dwarfs
\citep{aznar_cuadrado04}, pulsating stars \citep{silvester09}, hot subdwarfs
\citep{otoole05}, Ap and Bp stars \citep{wade00} and chemically peculiar stars
\citep{bagnulo02}.

\section{The weak field approximation}
The weak field approximation is an analytical solution to the radiative transfer
equation. The fundamental assumption is that the magnetic field vector is
constant and
its intensity is sufficiently
weak throughout the whole region of the atmosphere where an spectral line is
formed. Additionally, the line-of-sight velocity and any broadening mechanism 
have to be constant with height in the line formation region.
As a consequence, the magnetic field can be considered as a perturbation
to the zero-field case. Quantitatively, the approximation holds whenever
$\Delta \lambda_B\ll\Delta \lambda_D$, where $\Delta \lambda_B$ is the Zeeman
splitting and $\Delta \lambda_D$ is the dominant broadening mechanism (thermal,
rotation, etc.). From its definition, it is
clear that the weak field regime occurs at different field strengths
for different spectral lines (depending on the sensitivity to the magnetic 
field, the local temperature and the atomic mass) and for different
stellar objects (depending on any non-magnetic broadening mechanism for the
spectral lines, the rotation being the most efficient mechanism in cool stars). 

\begin{figure*}
\includegraphics[width=0.8\textwidth]{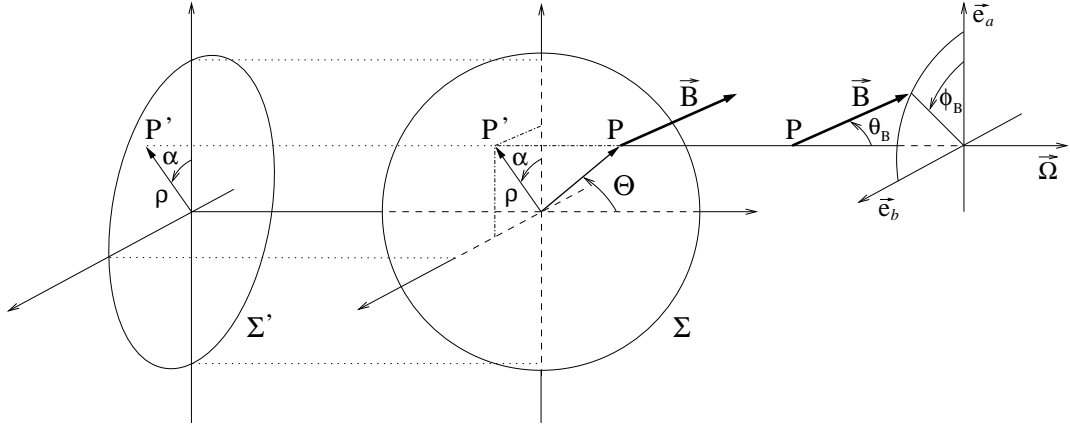}
\caption{Geometry of the stellar model. The symbol $\vec \Omega$ represents the
line of sight. The 
vector $\vec B$ displays the magnetic field vector in the stellar surface. Its
geometry is 
described in terms of the inclination with respect to the line of sight
$\theta_B$ and the 
azimuthal angle $\phi_B$ (the axis $\vec e_a$ being the 0 reference for the
azimuth). The 
position of a point P in the stellar surface ($\Sigma$) is defined only by the
astrocentric 
angle $\Theta$. Its projection (P') on the plane of the sky ($\Sigma'$) is
represented in 
cylindrical coordinates by the modulus $\rho$ and the angle $\alpha$, which is
referring to the $\vec e_a$ and $\vec e_b$ axis.}
\label{fig_geom}
\end{figure*}

To first order in $\Delta \lambda_B$, the intensity profile of a spectral line
formed in a
weak magnetic field is insensitive to the magnetic field. In other
words, it fulfils the transfer equation in the absence of a magnetic field. 
At this first order, the circular
polarisation profile, i. e., the Stokes
$V$ profile, for a given spectral line has the following expression:
\begin{equation}
V(x)=-C\Lambda g B_{||} \frac{\partial I(x)}{\partial x}.
\label{eq:v}
\end{equation}
The symbol $B_{||}$ stands for the longitudinal component of the magnetic field,
i. e., 
$B_{||}=B\cos{\theta_B}$, where $\theta_B$ is the inclination of the magnetic
field
with respect to the observer's line of sight ($\vec{\Omega}$; see Fig.
\ref{fig_geom}) and $B$ is the magnetic
field intensity. The symbol $g$ represents the
effective Land\'e factor of the line, which quantifies the magnetic sensitivity 
of the line. The effective Land\'e factor only depends
on the quantum numbers of the transition \citep[see,
e.g.,][]{landi_landolfi04}. 
Equation \ref{eq:v} is expressed in terms of the generic wavelength variable
$x$. If $x$ represents the wavelength $\lambda$, then the parameter $\Lambda$
equals the central wavelength of the line, $\lambda_0^2$.
However, it is customary in stellar spectropolarimetry to consider 
$x$ as the velocity in
Doppler units. In such a case, we have $\Lambda=c \lambda_0$, with $c$ the
speed
of light. The constant $C=4.67\times 10^{-13}$ G$^{-1}$ \AA$^{-1}$.

At first order in $\Delta \lambda_B$, both Stokes $Q$ and $U$ are zero. In order
to obtain an expression for the Stokes profiles characterizing linear
polarisation
we have to expand the radiative transfer equation to second order in $\Delta
\lambda_B$ and assume
that the spectral line is not saturated. Under these assumptions, the general
formulae
for the Stokes $Q$ and $U$ are:
\begin{align}
Q(x)&=-\frac{C^2}{4} \Lambda^2 G B_\perp^2 \cos{2\phi_B} \frac{\partial^2
I(x)}{\partial x^2}, \nonumber \\
U(x)&=-\frac{C^2}{4} \Lambda^2 G B_\perp^2 \sin{2\phi_B} \frac{\partial^2
I(x)}{\partial x^2}.
\label{eq:q_u}
\end{align}
The symbol $G$ plays the role of the effective Land\'e
factor for linear polarisation and quantifies the sensitivity of
linear polarisation to the magnetic field. Again, it is only a function 
of the quantum numbers of the transition \citep[see, e.g.,][]{landi_landolfi04}.
The 
symbol $B_{\perp}=B\sin{\theta_B}$ is the component of the magnetic field
perpendicular to the line of sight. The angle $\phi_B$ represents
the
azimuth of the magnetic field vector with respect to an arbitrary reference
direction ($\vec{e}_a$, $\vec{e}_b$ in Fig. \ref{fig_geom} show the coordinates
chosen for the 
reference of $Q > 0$).

When observing the polarised light in resolved sources like the Sun, we detect
specific intensities. However, when observing unresolved stars, the incoming
polarised
radiation is obtained as an integration in the plane of the sky of the
individual Stokes parameters at each point of the stellar surface. As a
consequence, 
the specific values of the Stokes flux vector depends on the surface
distribution of 
the magnetic field, on the centre-to-limb variation (CLV) of the radiation,
and the Doppler effect due to 
stellar rotation. In this paper we neglect rotation, the derived
expressions being valid for those cases where the broadening mechanisms 
dominate over rotation. In the Sun, since we observe local profiles, the main
broadening mechanism is typically thermal. In general, the polarimetric signals
of 99\% of 
the solar surface, the so-called quiet Sun, can be explained in terms of the
weak field regime 
in most spectral lines in the visible and the near-IR. For non-resolved objects,
the applicability depends 
strongly on the actual rotational velocity as compared with the temperature, the
observed spectral line, 
the spectral resolution, the magnetic field organisation, etc. The thermal
broadening 
depends on the square root of the ratio between the temperature and the 
atomic weight of the atom. Therefore, in general, the larger the temperature and
the 
lightest the atom, the larger the allowed rotational velocity. For low
resolution 
spectrographs ($R\sim 2000$), basically the only remaining spectral lines (that
are not wiped out
by the lack of spectral resolution) are from hydrogen. For these lines, for a
temperature 
of $T\sim 10000$ K, the maximum line of sight velocity is $10-20$ km s$^{-1}$.
However, when 
observing metal lines at spectral resolutions as high as $R\sim 60000$, the
allowed velocities 
are one order of magnitude lower.

It is customary to introduce a parametrised form of the CLV. In this work, we
assume
a quadratic form \citep[e.g.,][]{claret00,cox00}, which gives a good balance
between the quality of the CLV and the simplicity of the analytical expressions
presented in this paper for the polarised fluxes:
\begin{equation}
I(x, \mu)=I_0(x)\left(1-u-v+u\mu+v\mu^2\right) = I_0(x)f(\mu),
\label{eq:clv}
\end{equation}
where $I_0(x)$ is the intensity profile at disc centre ($\mu=1$). The 
parameters $u$ and $v$ have values between 0 and 1 and are supposed to
be constant along the spectral line (although they can vary from line to
line). Note that the values of $u$ and $v$ have to fulfil the condition 
that $I(x)>0$. The CLV is given in terms of $\mu=\cos \Theta$, where $\Theta$
is the astrocentric angle between the normal to a point in the stellar surface
and the line of sight. 
Using this law for the CLV (and neglecting rotation) the flux for Stokes $I$ is:
\begin{equation}
\mathcal{F}_I(x) = I_0(x) \int d\Sigma' f(\mu),
\end{equation}
where the integral is computed over the visible surface in the plane of the sky
$\Sigma'$. For an
arbitrary function $h(\rho,\alpha)$ of the polar coordinates of the 
stellar surface $\rho$ and $\alpha$, we have
\begin{equation}
\int \int d\Sigma' h(\rho,\alpha)= \int_0^1 \rho d\rho \int_0^{2\pi} d\alpha
h(\rho,\alpha).
\end{equation}
Note that the variable $\rho=\sin{\Theta}$, which means that
$\mu=\sqrt{1-\rho^2}$.
Plugging the quadratic expression considered for the CLV, the final closed
expression
for the integrated Stokes $I$ is found to be:
\begin{equation}
\mathcal{F}_I(x) = \pi I_0(x) \left( \frac{6-2u-3v}{6} \right).
\end{equation}
Integrating Eqs. (\ref{eq:v}) and (\ref{eq:q_u}) over the visible surface, we
end up
with the following expressions for the polarised flux:
\begin{align}
{\cal F}_V(x)&=-C \Lambda g \langle B_\parallel\rangle \frac{\partial {\cal
F}_I(x)}{\partial x}, \nonumber \\
{\cal F}_Q(x)&=-\frac{C^2}{4} \Lambda^2 G \langle B_\perp^2 \cos{2\phi_B}\rangle
\frac{\partial^2 {\cal F}_I(x)}{\partial x^2}, \nonumber \\
{\cal F}_U(x)&=-\frac{C^2}{4} \Lambda^2 G \langle B_\perp^2 \sin{2\phi_B}\rangle
\frac{\partial^2 {\cal F}_I(x)}{\partial x^2},
\label{eq:f_i}
\end{align}
with the definitions:
\begin{align}
\langle B_\parallel\rangle&=\frac{6}{\pi\left( 6-2u-3v \right)} \int d\Sigma' \
B_\parallel f(\mu) \nonumber \\
\langle B_\perp^2 \cos{2\phi_B}\rangle&=\frac{6}{\pi\left( 6-2u-3v \right)} \int
d\Sigma' \
B^2_\perp \cos{2\phi_B} f(\mu)\nonumber \\
\langle B_\perp^2 \sin{2\phi_B}\rangle&=\frac{6}{\pi\left( 6-2u-3v \right)} \int
d\Sigma' \
B^2_\perp \sin{2\phi_B} f(\mu).
\end{align}
Therefore, the weak-field approximation for the integrated polarised
flux remains formally the same but the components of the magnetic 
field $B_\parallel$ and $B_\perp$ appear weighted by the CLV law.
In general, the components of the magnetic field may depend on the 
position on the stellar surface in a complicated manner, so that the previous
integrals might not have closed expressions. In the simple case that the
magnetic field is
constant across the stellar surface, we recover
$\langle B_\parallel\rangle=B_\parallel$,
$\langle B_\perp^2 \cos{2\phi_B}\rangle= B_\perp^2 \cos{2\phi_B}$, and
$\langle B_\perp^2 \sin{2\phi_B}\rangle=B_\perp^2 \sin{2\phi_B}$.
One of the simplest non-trivial configuration we can consider
explicitly 
is that of a dipolar field. The magnetic field vector at each surface point
is then:
\begin{equation}
\mathbf{B}(\mathbf{r})=-\frac{H_d}{2}\left[\mathbf{e}-3(\mathbf{e}\cdot\mathbf{r
}
)\mathbf{r}\right],
\label{eq:dipolo}
\end{equation}
where the unitary vector $\mathbf{e}$ defines the orientation of the dipole and 
the unit vector $\mathbf{r}$ indicates positions on the stellar surface. The
quantity $H_d$ represents 
the magnetic field strength at the poles of the dipolar field. From Eq.
\ref{eq:f_i} 
and Eq. \ref{eq:dipolo} and after some algebra, it is possible to write the flux
of the
Stokes 
parameters as:
\begin{align}
{\cal F}_V(x)&=-\frac{1}{10}C\Lambda g
\frac{15+u}{6-2u-3v}H_\parallel \frac{\partial {\cal F}_I(x)}{\partial x}
\nonumber \\
{\cal F}_Q(x)&=-\frac{C^2}{4480}\Lambda^2 G
\frac{420-68u-105v}{6-2u-3v} H_\perp^2 \cos{2\phi_d} \frac{\partial^2
{\cal F}_I(x)}{\partial x^2} \nonumber \\
{\cal F}_U(x)&=-\frac{C^2}{4480}\Lambda^2 G
\frac{420-68u-105v}{6-2u-3v} H_\perp^2 \sin{2\phi_d} \frac{\partial^2
{\cal F}_I(x)}{\partial x^2},
\label{eq:stokes_dipole}
\end{align}
where $H_\parallel=H_d\cos{\theta_d}$ and $H_\perp=H_d\sin{\theta_d}$,
$\theta_d$ 
being the inclination of the axis of the dipole while $\phi_d$
is the azimuth of the dipole \citep[for a similar derivation see][]{landolfi93}.

The previous formalism has demonstrated that the weak-field approximation leads
to formally the same expressions either if we consider specific intensities
(that
is applied whenever spatial resolution is available) or if we consider fluxes
(whenever the object of interest cannot be resolved). In other words, the
observed 
circular (linear) polarised spectrum is
proportional to the first (second) derivative of the observed intensity through
the longitudinal (orthogonal) component of the magnetic field. The only
difference resides on 
the exact definition of the components of the field which, in the spatially
unresolved case, are an average over the stellar surface weighted by the CLV.
Therefore, for the sake of simplicity, it is possible to combine
all expressions on the following general ones:
\begin{align}
{\cal V}&=-C {\cal B}_\parallel  {\cal I'} \nonumber \\
{\cal Q}&=-C^2 {\cal B}^2_\perp \cos{2\phi} {\cal I''} \nonumber \\
{\cal U}&=-C^2 {\cal B}^2_\perp \sin{2\phi} {\cal I''}.
\label{eq:generales}
\end{align}
The meaning of the newly defined variables is summarized in Tab.
\ref{tab:variables}.

\begin{table}
\caption{Variables in Eq. \ref{eq:generales}.}
\label{tab:variables}
\centering{\begin{tabular}{c|cc}
\hline
Variable & Resolved & Dipole \\
\hline
$\mathcal{I}$ & $I$ & $\mathcal{F}_I$ \\
$\mathcal{V}$ & $V$ & $\mathcal{F}_V$ \\
$\mathcal{Q}$ & $Q$ & $\mathcal{F}_Q$ \\
$\mathcal{U}$ & $U$ & $\mathcal{F}_U$ \\
$\mathcal{I'}$ & $\Lambda g\frac{\partial I(x)}{\partial x}$ & 
$\frac{1}{10}\frac{15+u}{6-2u-3v}\Lambda g\frac{\partial{\cal F}_I(x)}{\partial
x}$ \\
$\mathcal{I''}$ & $\frac{1}{4}\Lambda^2 G \frac{\partial^2 I(x)}{\partial
x^2}$ & 
$\frac{1}{4480}\frac{420-68u-105v}{6-2u-3v}\Lambda^2 G \frac{\partial^2{\cal
F}_I(x)}{\partial x^2}$ \\
${\cal B}_\parallel$ & $B_\parallel$ & $H_\parallel$ \\
${\cal B}_\perp$ & $B_\perp$ & $H_\perp$ \\
$\phi$ & $\phi_B$ & $\phi_d$ \\
$\theta$ & $\theta_B$ & $\theta_d$ \\
${\cal B}$ & $B$ & $H_d$ \\
\hline
\end{tabular}}
\end{table}

\section{Estimation of the magnetic field vector}
Once the model is set, our aim is to infer the magnetic field vector 
parametrised in terms of ${\cal B}_\parallel$, ${\cal B}_\perp$, and $\phi$
from 
observations of a set of spectral lines
$(\mathcal{I}^i,\mathcal{Q}^i,\mathcal{U}^i,\mathcal{V}^i)$
with $i=1\ldots n_\mathrm{lines}$. Assuming that the weak field approximation
can be 
applied for the set of observed spectral lines and that observations are
corrupted
with uncorrelated Gaussian noise, we can use a least-squares estimator (maximum
likelihood)
to retrieve the magnetic field vector (see the Appendix for more details). The
$\chi^2$ merit function is defined as:
\begin{align}
\chi^2 = &\sum_{ij} \frac{({\cal V}^i_j-{\cal
V}^{i;mod}_j)^2}{(\sigma^i_{{\cal
V}j})^2}+\sum_{ij} \frac{({\cal Q}^i_j-{\cal
Q}^{i;mod}_j)^2}{(\sigma^i_{{\cal Q}j})^2} \nonumber \\
+&\sum_{ij} \frac{({\cal U}^i_j-{\cal
U}^{i;mod}_j)^2}{(\sigma^i_{{\cal U}j})^2},
\label{eq:chi2}
\end{align}
where the subindex $j$ indicates the position along the 
coordinate $x$. The label $mod$ refer to the model for Stokes profiles, and 
the superindex $i$ labels the spectral lines. The previous merit
functions consider the quite general case in which the standard deviation of the
corrupting Gaussian noise is different for Stokes $Q$, $U$, and $V$ and for each
wavelength
point $x_j$. However, in practice, we have the simpler situation in which 
$\sigma^i_{{\cal V}j}\approx\sigma^i_{{\cal Q}j}\approx\sigma^i_{{\cal
U}j}=\sigma^i_j$. 
Moreover, although the number of photons arriving in the line cores are smaller
than in the far wings,
we make the approximation that the noise variance is wavelength-independent.
Furthermore,
we also consider that it is independent of the considered line, so that
$\sigma^i_j=\sigma$. These simplifications
lead to less cluttered expressions for the inferred parameters. However, we point
out that
the general expressions that emerge from the optimization of Eq. (\ref{eq:chi2})
can 
be found in the Appendix.

In order to infer a certain parameter, we have to find the global minimum of the
$\chi^2$. This
is trivially obtained by solving the non-linear system of equations obtained by
setting
the derivatives of the $\chi^2$ function with respect to that parameter to
zero. By
so doing (see Appendix), we can obtain the expression of the magnetic field
vector in terms of the observables:
\begin{align}
{\cal B}_{||}&=-\frac{1}{C}\frac{\sum_{ij}{\cal V}^i_j {\cal
I'}^i_j}{\sum_{ij}({\cal
I'}^i_j)^2} \nonumber \\
{\cal B}_{\perp}^2&=\frac{1}{C^2}\frac{\sqrt{(\sum_{ij}{\cal Q}^i_j {\cal
I''}^i_j)^2+(\sum_{ij}{\cal U}^i_j {\cal I''}^i_j)^2}}{\sum_{ij}({\cal
I''}^i_j)^2} \nonumber \\
\phi&=\frac{1}{2} \arctan{\frac{\sum_{ij}{\cal U}^i_j {\cal
I''}^i_j}{\sum_{ij}{\cal Q}^i_j {\cal I''}^i_j}}+\phi_0,
\label{eq:inference}
\end{align}

and the derived quantities:
\begin{align}
\tan{\theta}&=\frac{\cal B_\perp}{\cal B_{||}} \nonumber \\
{\cal B}&=\sqrt{{\cal B}_\parallel^2+{\cal B}_\perp^2}.
\label{eq:inference2}
\end{align}
The phase shift $\phi_0$ is used to set the correct quadrant
for the azimuth and depends on the sign of the numerator
\hbox{$\mathfrak{U}=\sum_{ij}{\cal U}^i_j {\cal
I''}^i_j$} and the denominator \hbox{$\mathfrak{Q}=\sum_{ij}{\cal Q}^i_j {\cal
I''}^i_j$} as follows. If $\mathfrak{Q}\ne 0$ then:
\begin{align}
\phi_0=
\begin{cases}
0 & \text{if $\mathfrak{U} \ge 0$ and $\mathfrak{Q} > 0$} \\
\pi & \text{if $\mathfrak{U} < 0$ and $\mathfrak{Q} > 0$} \\
\frac{\pi}{2} & \text{if $\mathfrak{Q} < 0$}
\end{cases}
\end{align}
In the case $\mathfrak{Q}=0$ then:
\begin{align}
\phi_0=
\begin{cases}
\frac{\pi}{4} & \text{if $\mathfrak{U} > 0$} \\
\frac{3\pi}{4} & \text{if $\mathfrak{U} < 0$}
\end{cases}
\end{align}
If both ${\mathfrak Q}$ and ${\mathfrak U}$ are zero, the angle $\phi$ is
obviously
undefined.

\begin{figure*}
\includegraphics[width=0.45\textwidth]{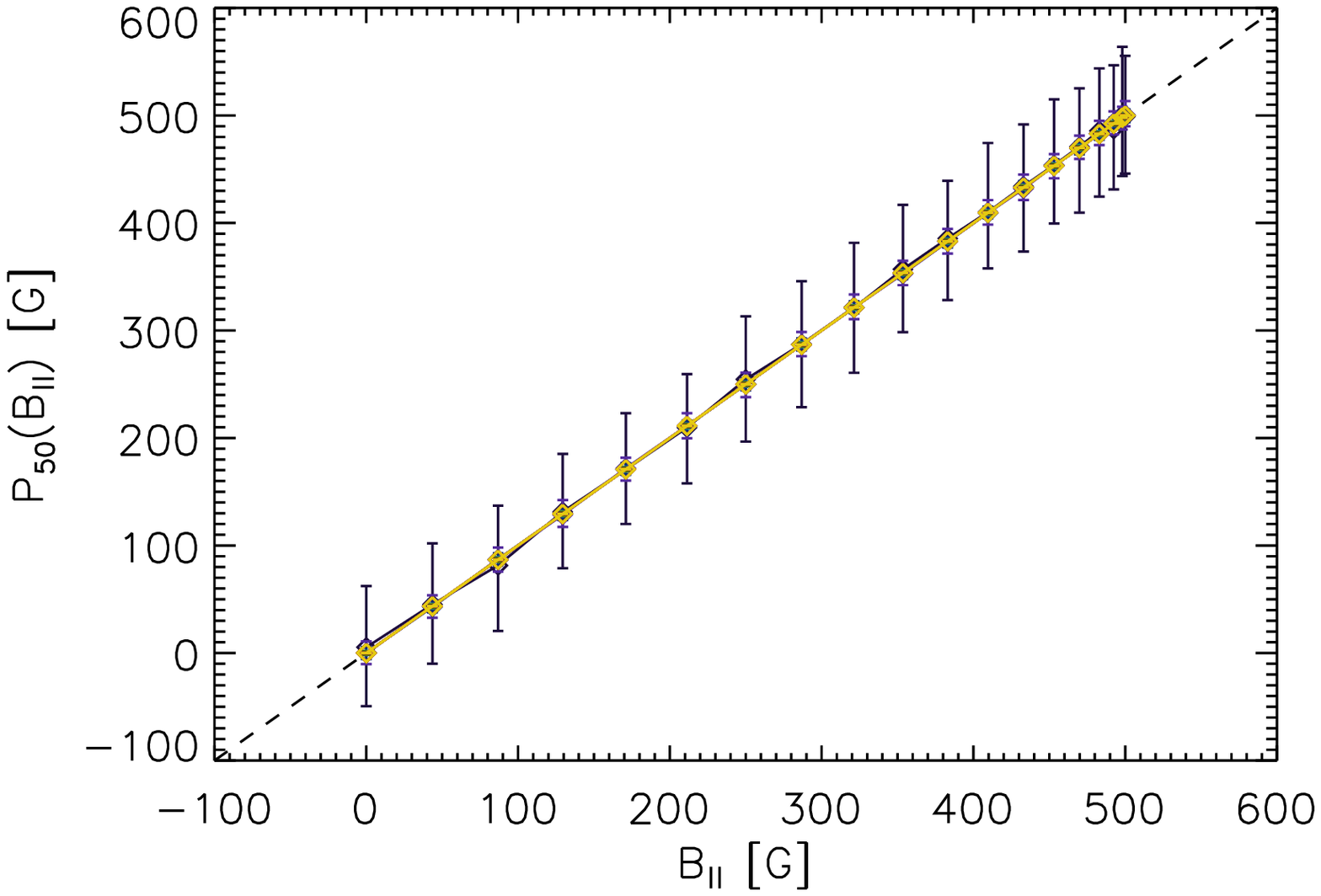}
\includegraphics[width=0.45\textwidth]{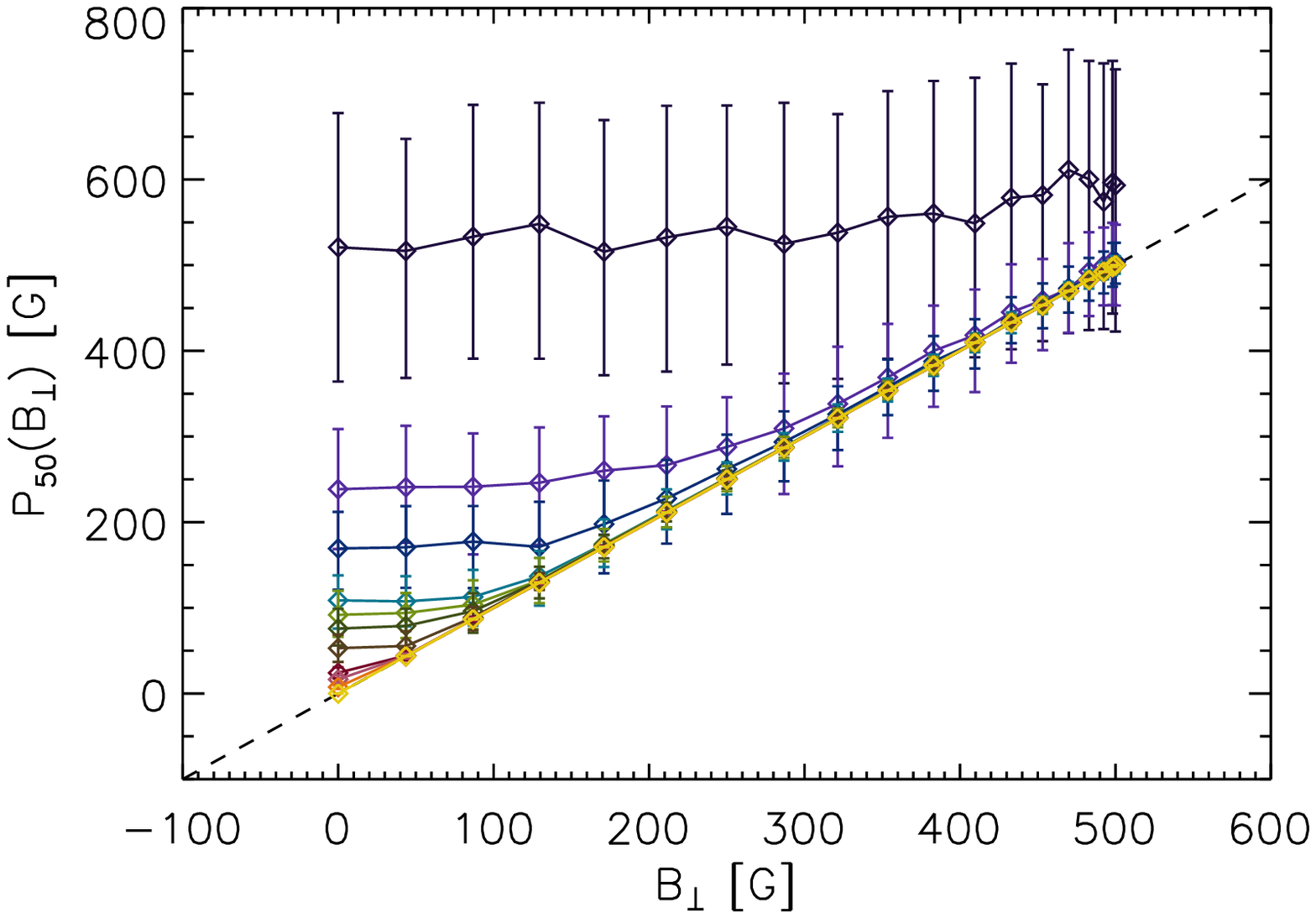}\\
\includegraphics[width=0.45\textwidth]{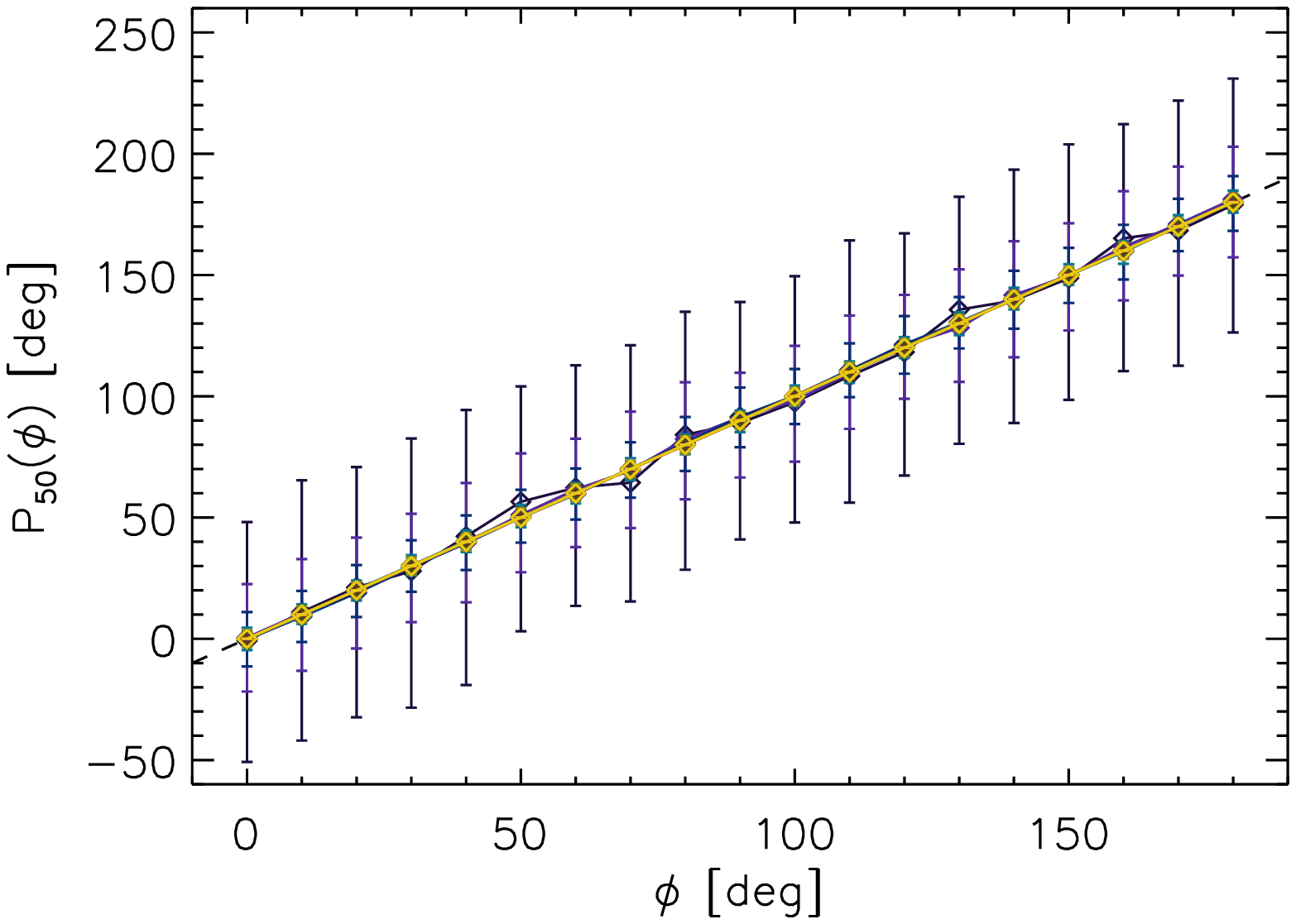}
\includegraphics[width=0.45\textwidth]{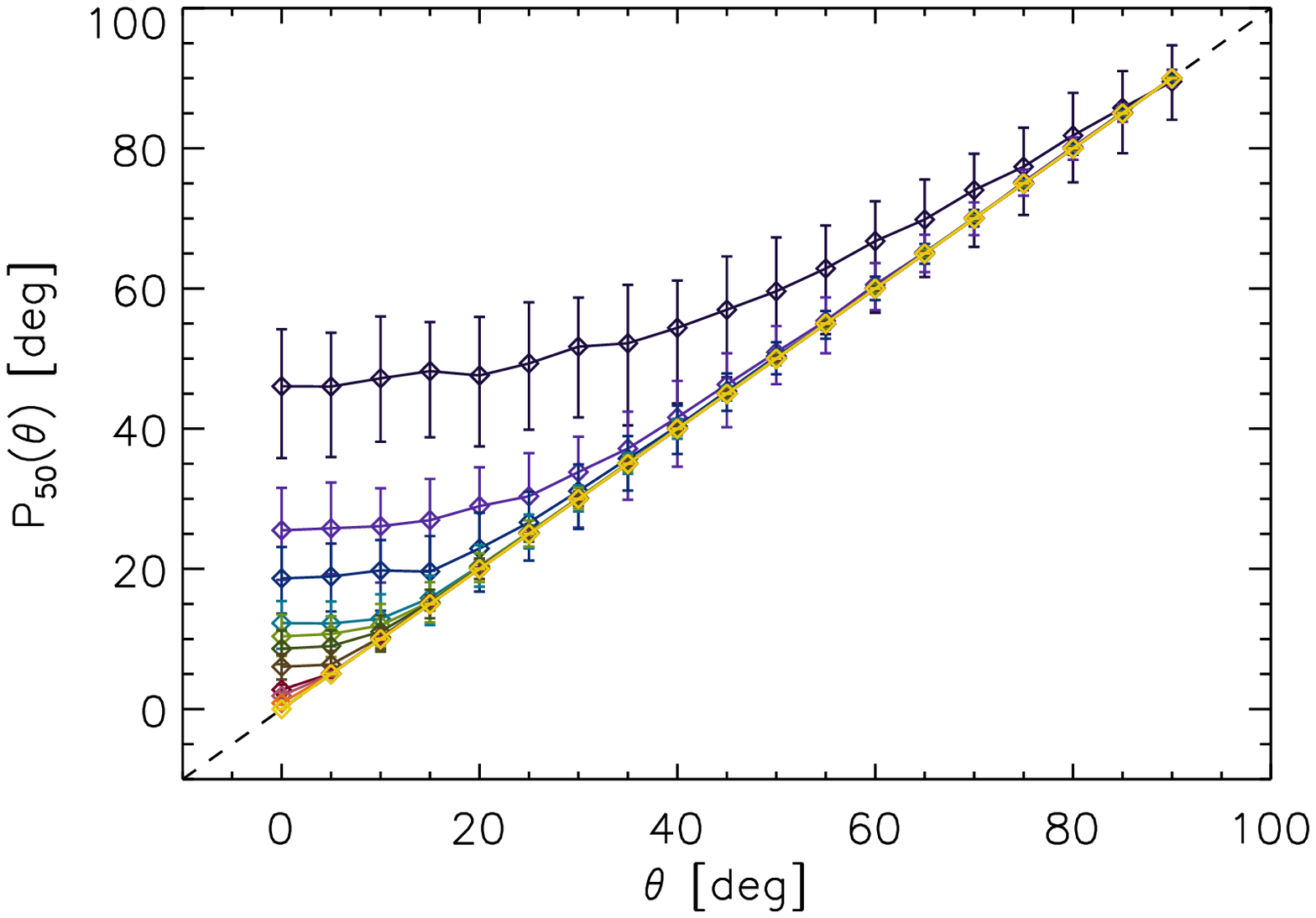}
\caption{Bias of the maximum likelihood estimator of the magnetic field in the
weak field approximation. Different color lines represent different values of
the noise level added to the synthetic profiles of the Fe\,{\sc i} line at
5250.2 \AA\ . The color code is the following: yellow represents the inversion
of the profiles without noise. Orange displays the results for a noise level of
$10^{-5}$ I$_\mathrm{c}$, pink for $5\times 10^{-5}$ I$_\mathrm{c}$, red for
$10^{-4}$ I$_\mathrm{c}$, brown for $5\times 10^{-4}$ I$_\mathrm{c}$, dark green
for $10^{-3}$ I$_\mathrm{c}$, light green for $1.5\times 10^{-3}$
I$_\mathrm{c}$, light blue for $2\times 10^{-3}$ I$_\mathrm{c}$, dark blue for
$5\times 10^{-3}$ I$_\mathrm{c}$, light violet for $10^{-2}$ I$_\mathrm{c}$, and
dark violet for $5\times 10^{-2}$ I$_\mathrm{c}$. The I$_\mathrm{c}$ is the
continuum intensity.\label{fig:bias_sigma_cte}}
\end{figure*}

The estimated errors can be computed from the covariance matrix
\citep[e.g.,][]{numerical_recipes86}. 
In the simple case we consider of equal wavelength-independent standard
deviation for all 
spectral lines, the covariance matrix is diagonal (no correlation between the
parameters), 
and the errors at a confidence level of 68.3\% (one sigma) are expressed as:
\begin{align}
\delta {\cal B}_{||}&=\pm \frac{\sigma}{C\sqrt{\sum_{ij}({\cal
I'}^i_j)
^2}} \nonumber \\
\delta {\cal B}_{\perp}&=\pm \frac{\sigma}{2C^2 {\cal
B}_{\perp}\sqrt{\sum_{ij}({\cal I''}^i_j) ^2}} \nonumber \\
\delta \phi&=\pm \frac{\sigma}{2C^2 {\cal B}^2_{\perp}\sqrt{\sum_{ij}({\cal
I''}^i_j) ^2}} \nonumber \\
\delta \theta&=\pm \frac{\sqrt{{\cal B}_\perp^2\delta {\cal B}_\parallel^2 +
{\cal B}_\parallel^2\delta {\cal B}_\perp^2}}{{\cal B}_\parallel^2+{\cal
B}_\perp^2} \nonumber \\
\delta {\cal B}&=\pm \sqrt{\frac{{\cal B}_\parallel^2\delta {\cal B}_\parallel^2
+ {\cal B}_\perp^2\delta {\cal B}_\perp^2}{{\cal B}_\parallel^2+{\cal
B}_\perp^2}},
\label{eq:error_bars}
\end{align}

\section{Bias of the maximum likelihood estimator}
It is widely known that all maximum likelihood estimators may suffer
from biases. The bias is the difference between the value of the estimator 
and the true value of the parameter. Each individual parameter has to be studied
separately using analytical/numerical simulations to understand to which extent 
the estimation of ${\cal B}_\parallel$, ${\cal B}_\perp$, and $\phi$ are
subjected to bias.
For simplicity
the simulations we present in the following refer to the resolved case in which
we
use specific intensities. However, the behaviour of the estimator is the
same also for the stellar case.

In order to carry out the simulations, we focus on the Fe \textsc{i} line with
central
wavelength \hbox{$\lambda_0=5250.2$ \AA}. This spectral line is produced by the
transition $^5D_0 - ^7D_1$. 
The values of the effective Land\'e factor for circular polarisation is $g=3$,
while it is $G=9$ for linear polarisation. We consider a constant magnetic
field 
strength of $B=500$ G, which is sufficiently weak so that the weak-field
approximation can still be considered. The inclination and the
azimuth of the field are set to vary uniformly between $0$ and $180^\circ$.
Letting
the azimuth vary in this interval, we avoid the $180^\circ$ ambiguity in the
azimuth present
in the radiative transfer equation. For simplification, we consider
a Gaussian intensity profile in the form:
\begin{equation}
I(\lambda)=1-d_c \exp \left[ - \frac{(\lambda-\lambda_0)^2}{2 w^2} \right],
\end{equation}
where $d_c=1/2$ and $w=0.05$ \AA .
The parameters for the Gaussian profile have been fixed to fit a solar
observation
obtained with the IMaX instrument. From the IMaX observational capabilities, it
can be verified that the \hbox{$5250.2$ \AA} line is in the weak-field regime up
to $\sim 1$ kG (with the line assumed to be in local thermodynamical
equilibrium in a quiet Sun model atmosphere). 
We have synthesized 500 profiles for different combinations of the inclination
and azimuth 
and different noise realisations. The noise added has a Gaussian statistics and
we
consider the effect of different standard deviations. We use Eqs.
(\ref{eq:inference})
to compute the inferred values of the parameters. Since we repeat the experiment
for different realisations of the 
noise, we end up with a distribution of values for each parameter. We adopt the
median value as the estimation of the parameter (the percentile 50 $P_{50}$, i.
e. , the value of the parameter that contains 50\% of the area of the
distribution). To quantify the dispersion
produced by noise we use the percentiles 16 and 84 (which encompass one
standard deviation around the estimated value).

Figure \ref{fig:bias_sigma_cte} displays the inversion of the synthetic profiles
with different values of the standard deviation of the noise, from $0$ to
$5\times 10^{-2}$ in
units of the continuum intensity, I$_\mathrm{c}$. It is clear from the
upper and lower left panels that both the longitudinal component of
the magnetic field and the azimuthal angle are unbiased quantities. This means
that the estimated value statistically coincides with the original one. As
expected, the
dispersion grows with the increasing noise level. 

Contrary to $B_\parallel$ and $\phi_B$ (note that the notation for the magnetic
field vector
is the one associated to resolved sources), the transversal
component of the magnetic field and hence the inclination angle presents a
non-zero bias. Except for the case in which there is no added noise, small
transversal
components of the field (smaller or at the level of the noise amplitude)
are overestimated. The fundamental reason is that the expression for 
$B_\perp$ in Eq. (\ref{eq:inference})
is not robust against noise. It can be verified that if $\mathcal{Q}^i_j$
and $\mathcal{U}^i_j$ are at the noise level (can be described as Gaussian
distributions with zero mean and variance $\sigma^2$), $B_\perp$ is a random
variable following the probability distribution:
\begin{align}
p(B_\perp) = \frac{2 C^4 \left( \sum_{ij}({\cal I''}^i_j)^2 \right)}{\sigma^2}
B_\perp^3
\exp \left[ - \frac{C^4 \sum_{ij}({\cal I''}^i_j)^2 B_\perp^4}{2\sigma^2}
\right].
\end{align}
The value of this distribution for a given percentile $c$ fulfils:
\begin{align}
B_\perp^c = \left( \frac{-2 \ln(1-c)}{\sum_{ij}({\cal I''}^i_j)^2}
\right)^{1/4}\frac{\sqrt{\sigma}}{C}.
\label{eq:bias}
\end{align}
The percentile 50 ($c=0.5$) correctly captures the value of the bias at small
values
of $B_\perp$. Once this value is computed, if the inferred value 
of $B_\perp$ is similar to this value, one should be aware that
the correct value of $B_\perp$ might be smaller. If this is not
taken into account, the estimated inclination of the field is larger
than the real one and artificially horizontal fields might be inferred. 
Note also that in the stellar case, the inferred inclination of the dipole axis 
would be more inclined than the real one. 

\section{The particular case of the Least-Squares-Deconvolution profile}
Most detections of faint signals in stellar atmospheres have been
evidenced by adding many spectral lines. The polarimetric signal per
spectral resolution element is known to be well below the noise level,
so that line addition is a must to fight against photon noise.
The most widely used and successful
technique that combines the information of many spectral lines is the
Least-Squares
Deconvolution (LSD) technique \citep{donati97}. The equations presented in this
paper allow us to retrieve the magnetic field vector from the polarised stellar
spectra taking into account many spectral lines. Thanks to it, we can rewrite 
the equations to directly extract information from the LSD profiles. 

The LSD technique is fundamentally based on the application of the 
weak-field approximation and on the assumption of a constant CLV for all 
spectral lines (in our case, $u$ and $v$ are set constant). This means that all
Stokes profiles for each spectral line can be computed with a single spectral
profile 
whose proportionality constant changes from line to line
\citep[see][]{donati97}:
\begin{align}
{\cal I}^i&=\eta^i \bar{{\cal I}}\\
{\cal V}^i&=\eta^i \Lambda^i g^i \bar{{\cal V}}\\
{\cal Q}^i&=\eta^i (\Lambda^i)^2 G^i \bar{{\cal Q}}\\
{\cal U}^i&=\eta^i (\Lambda^i)^2 G^i \bar{{\cal U}},
\end{align}
where $\eta^i$ is the line depth and $\bar{{\cal I}}$, $\bar{{\cal Q}}$, 
$\bar{{\cal U}}$, and $\bar{{\cal V}}$ are the LSD profiles that can be
computed 
using a least-squares procedure as explained in \cite{donati97}.
\textbf{We refer to
\cite{kochukhov10} for an in-depth analysis of the assumptions and potential
problems
of LSD.}
From the previous equations, it is easy to show that it is possible
to infer the magnetic field vector using:
\begin{align}
{\cal B}_\parallel&=-\frac{1}{CK}\frac{\sum_j \bar{{\cal V}}_j \frac{\partial
\bar{{\cal
I}}_j}{\partial x}}{\sum_j \left(\frac{\partial \bar{{\cal I}}_j}{\partial
x}\right)^2} \nonumber \\
{\cal B}_\perp&=\frac{1}{C^2K'}\frac{\sqrt{\left(\sum_j \bar{{\cal Q}_j}
\frac{\partial^2 \bar{\cal
I}_j}{\partial x^2}\right)^2+\left(\sum_j \bar{{\cal U}_j} \frac{\partial^2
\bar{\cal I}_j}{\partial x^2} \right)^2}}{\sum \left(\frac{\partial^2 \bar{{\cal
I}}_j}{\partial x^2}\right)^2} \nonumber \\
\phi&=\frac{1}{2}\arctan{\frac{\sum_j \bar{{\cal U}}_j \frac{\partial^2
\bar{{\cal I}}_j}{\partial x^2}}{\sum_j \bar{{\cal Q}}_j \frac{\partial^2
\bar{{\cal I}}_j}{\partial x^2}}}+\phi_0,
\end{align}
where $K=1$ and $K'=1/4$ for the case of an unresolved star with constant
magnetic field 
and 
\begin{equation}
K =\frac{1}{10}\frac{15+u}{6-2u-3v}, \qquad
K' =\frac{1}{4480}\frac{420-68u-105v}{6-2u-3v}
\end{equation}
for the stellar dipole. Note that
it is not necessary to assume averaged atomic parameters for the LSD profile,
treating it as a {\it mean} spectral line.
In our case, the estimated magnetic field vector only depends on the
observables, the atomic parameters of each observed spectral line (which is
necessary to compute the LSD profile), and the assumed CLV coefficients.

\begin{figure*}
\includegraphics[width=0.45\textwidth, bb= 0 54 489 337]{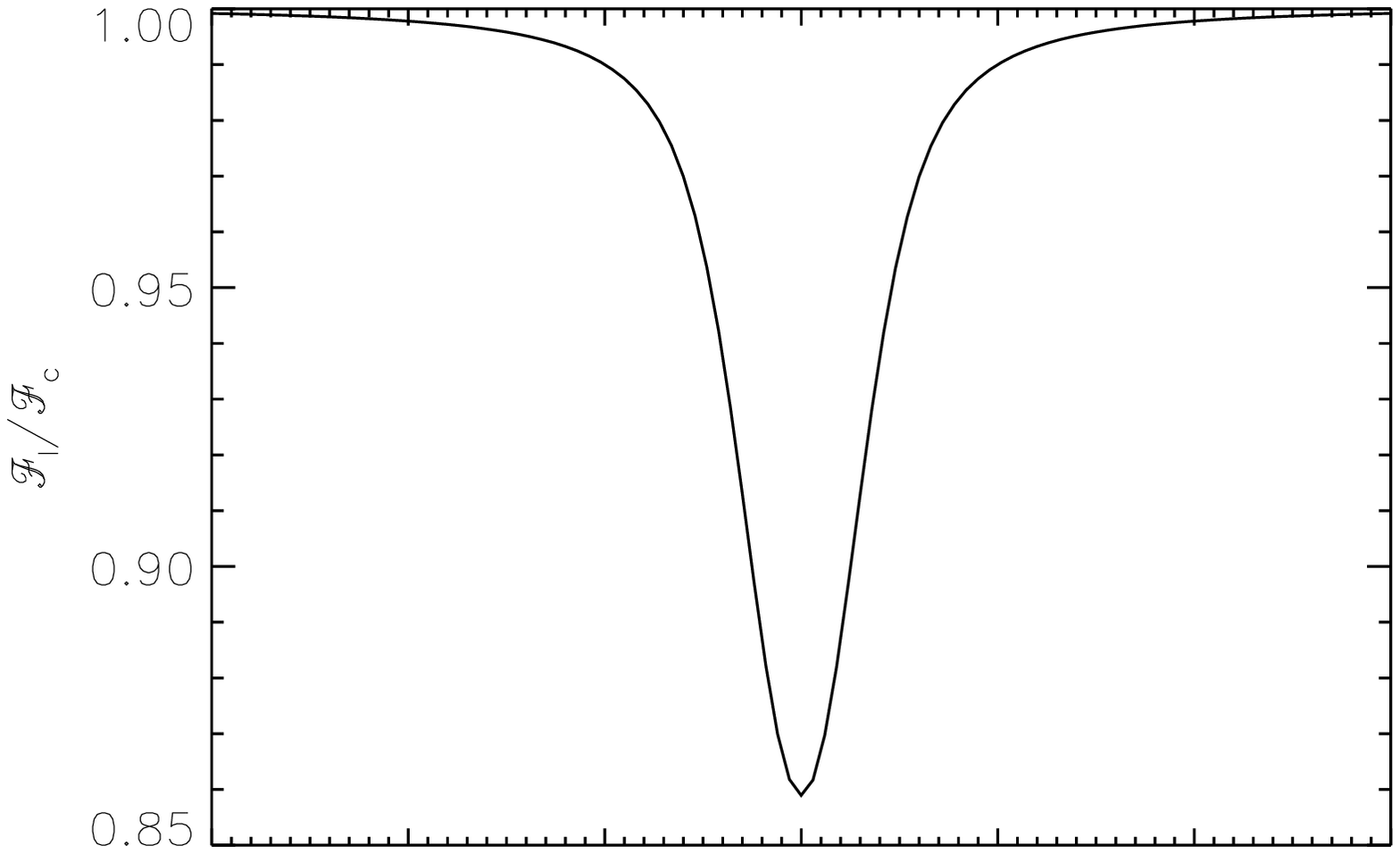}
\includegraphics[width=0.45\textwidth, bb= 0 54 489 337]{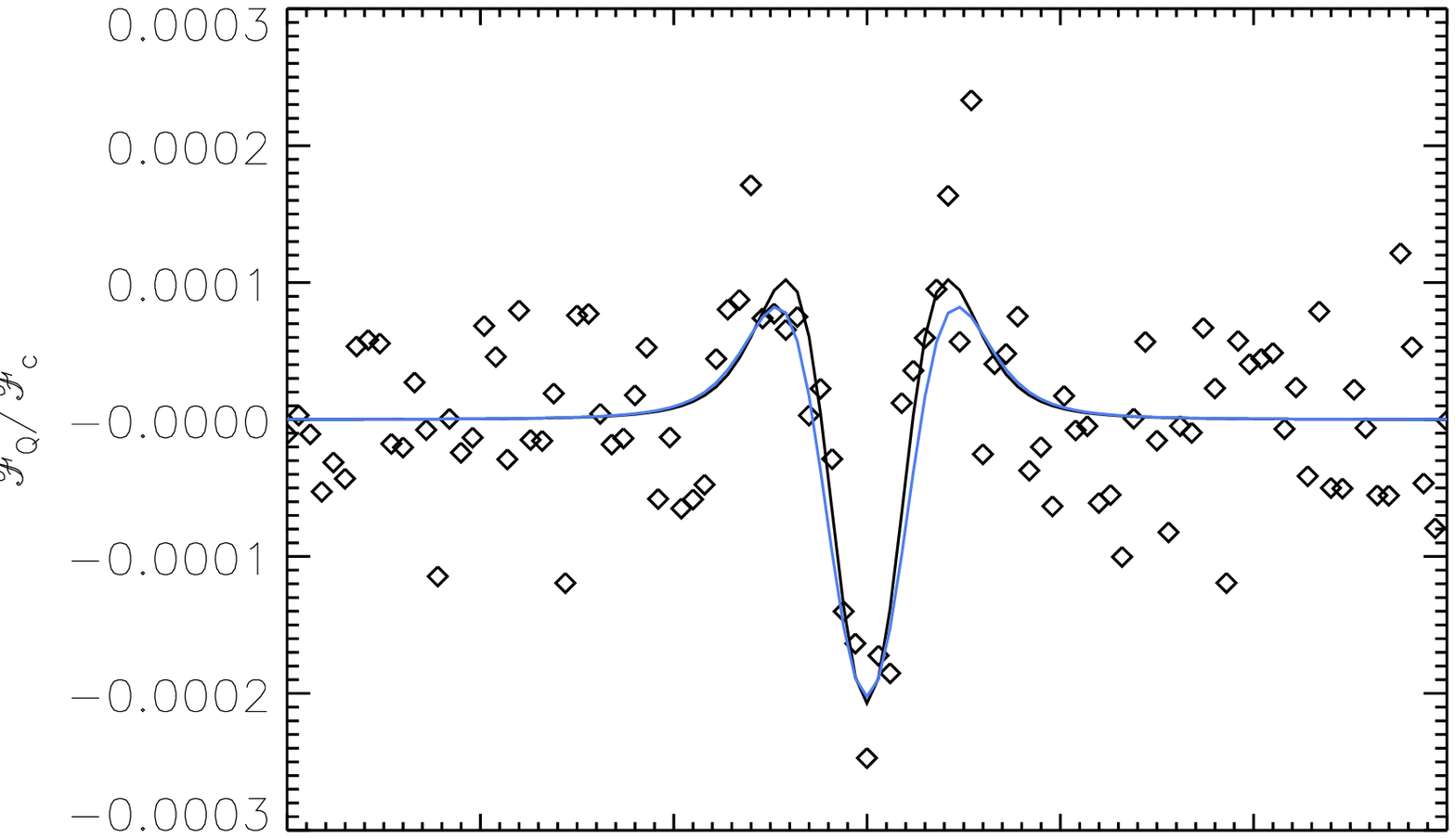}\\
\includegraphics[width=0.45\textwidth, bb= 0 0 489 337]{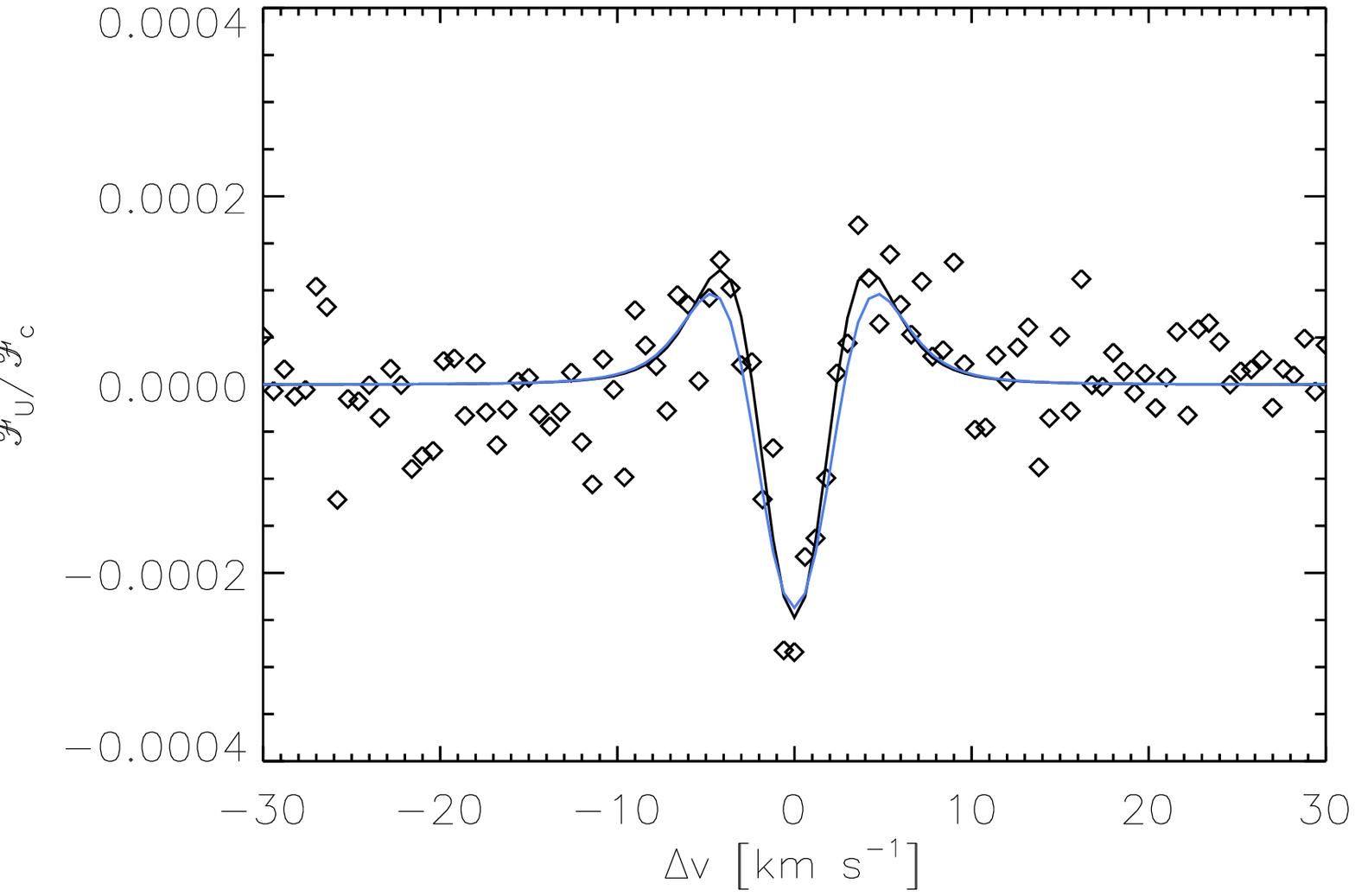}
\includegraphics[width=0.45\textwidth, bb= 0 0 489 337]{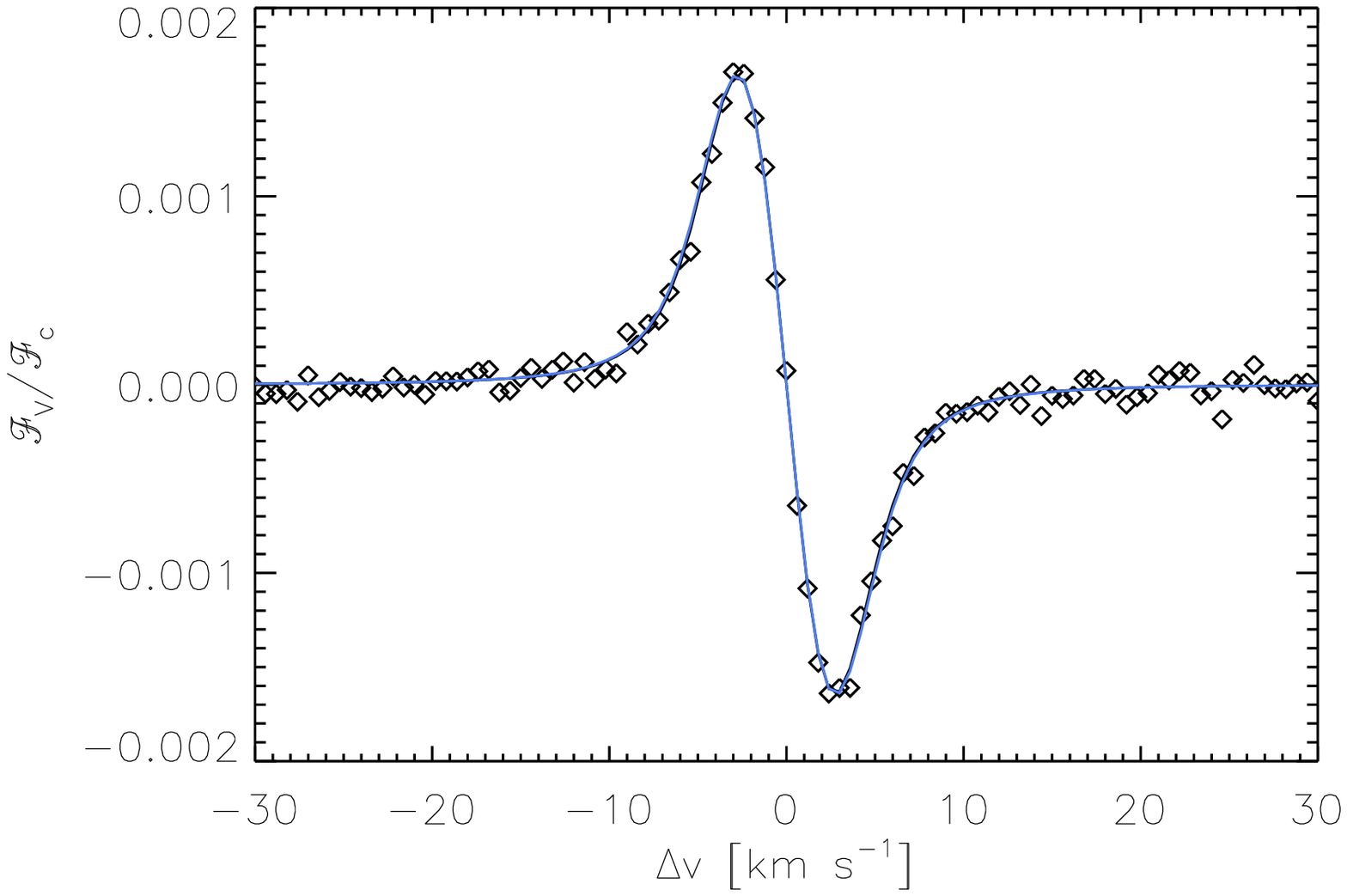}
\caption{Inversion of a synthetic stellar dipole. The black lines display the
synthetic flux of a dipole with a strength $H_d$=1500 G, an inclination 
$\theta_d=80^\circ$ with respect to the line of sight and an azimth of
$\theta_d=25^\circ$. The rhombs are the synthetic polarised flux including an
additive Gaussian noise with a standard deviation of $5\times 10^{-5}$ ${\cal
F}_c$. The blue lines are the fluxes inferred from the inversion using Eq. 13.}
\label{fig:test_dipole}
\end{figure*}

\section{Illustrative examples}

The inference power of the expressions developed in the previous sections are illustrated with the
aid of two different examples. The first one consists of a simulated
stellar dipole using a Milne-Eddington \citep[e.g.,][]{landi_landolfi04} atmosphere and the second one is a
particularly interesting observational example in which we can illustrate the
effect of the bias in the transverse component of the magnetic field with
spatial resolution. 

Considering the stellar dipole, we simulate a Milne-Eddington atmosphere with a
source function that varies as:
\begin{equation}
S(\tau) = S_0 (1+\beta \tau).
\end{equation}
For simplicity, we choose $\beta=1$ and consider a static atmosphere at all
points of the stellar surface. We assume a spectral line centred at
$\lambda=5000$ \AA\ with a Doppler broadening of 0.04 \AA\ . Both the circular and linear effective
Land\'e factors are equal to 1. The CLV variation of the Milne-Eddington 
atmosphere (that we have forced to be wavelength independent) gives $u=2/3$. We integrate the 
Stokes signals on the visible stellar surface and we add Gaussian noise to the final integrated flux
with a standard deviation of $5\times 10^{-5}$ in units of the intensity flux at
the continuum, ${\cal F}_c$. 

Figure \ref{fig:test_dipole} shows the flux of the simulated Stokes parameters
coming from a dipole field with the following parameters: $H_d=1500$ G,
$\theta_d=80^\circ$, and $\phi_d=25^\circ$. The black lines represent the
synthetic fluxes, and the rhombs the synthetic observations with added noise. 
We invert the noisy Stokes fluxes using
Eqs. \ref{eq:inference}, \ref{eq:inference2} and \ref{eq:error_bars} and obtain $H_{||}=265.5\pm 2.4$ G, $H_\perp=1501.6\pm 25.7$ G,
$\phi_d=27.2^\circ\pm 2.0$, and the derived quantities $H_d=1524.9\pm 25.3$ G
and $\theta_d=80.0^\circ \pm 0.2$. All quantities are nicely 
recovered. In fact, the bias estimations for the perpendicular component
$H_\perp$ for a the percentiles 16, 50, and 84 (i. e. containing one sigma
probability) are 302.0, 426.5, and 543.8 G, respectively. This means that the
computed $H_\perp$ is reliable. Now, we assume a weak dipole with $H_d=100$ G,
$\theta_d=20^\circ$, and $\phi_d=25^\circ$, and the same noise level. The
inferred parameters are $H_{||}=99.7\pm 2.3$ G, $H_\perp=333.9\pm 109.0$ G,
$\phi_d=39.8^\circ\pm 37.4$, $H_d=348.3 \pm 104.5$ G, and
$\theta_d=73.4^\circ\pm 5.1$. In principle, both the longitudinal and the
transverse components (as well as the inclination) should be well recovered but the
azimuth remains undetermined. However, the bias of $H_\perp$ for the
percentiles 16, 50, and 84 are 293.2, 414.1, and 528.0 G, respectively. Therefore, the
inferred perpendicular component of the dipole is consistent with a bias. This
implies that the inferred value, although having a small (Gaussian) error, it has to
be considered an upper limit. In this case, we know that the perpendicular
component of the dipole is very small (34 G). Consequently, we overestimate both the
perpendicular component and the intrinsic strength of the dipole. Additionally, the
dipole appears to be much more inclined that in reality.

\begin{figure*}
\includegraphics[width=0.33\textwidth, bb= 60 40 406 341]{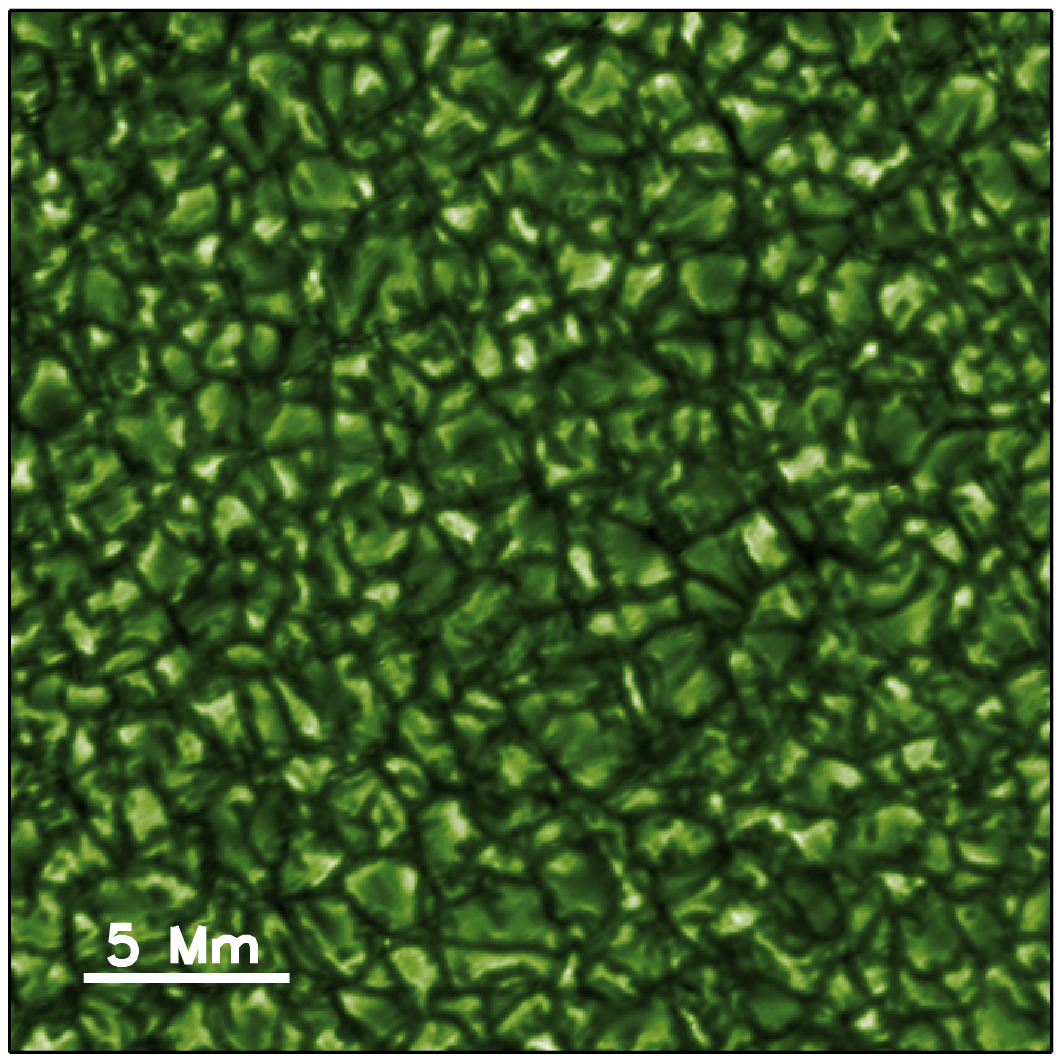}
\includegraphics[width=0.33\textwidth, bb= 60 40 406 341]{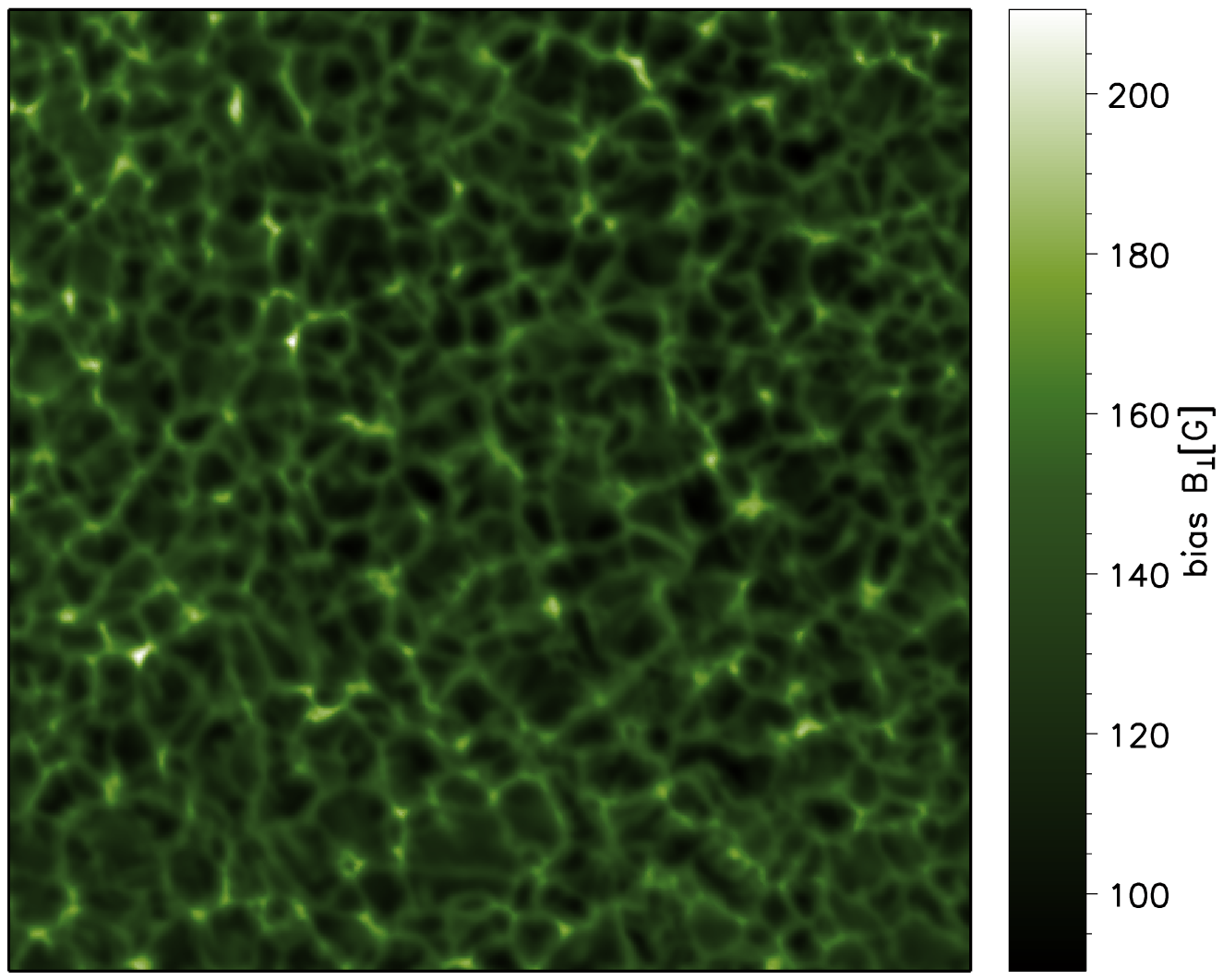}
\caption{Left: continuum intensity at 227 m\AA\ from the core of the 5250.2 \AA.
 Right: Median value of the bias for the transverse component of the magnetic
field computed with Eq. \ref{eq:bias} and $c=0.5$.}
\label{fig:imax1}
\end{figure*}

\begin{figure*}
\includegraphics[width=0.33\textwidth, bb= 60 40 406 341]{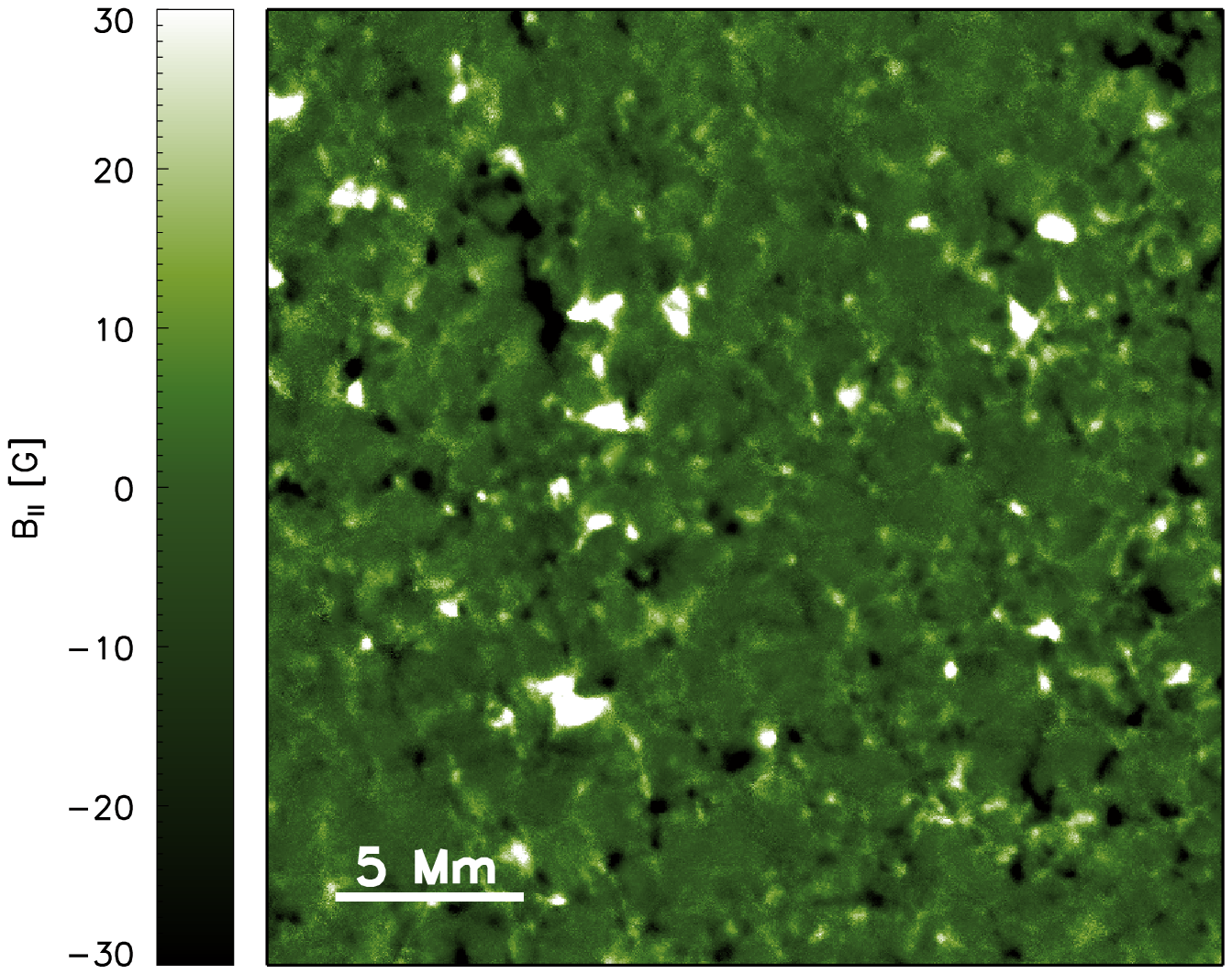}
\includegraphics[width=0.33\textwidth, bb= 60 40 406 341]{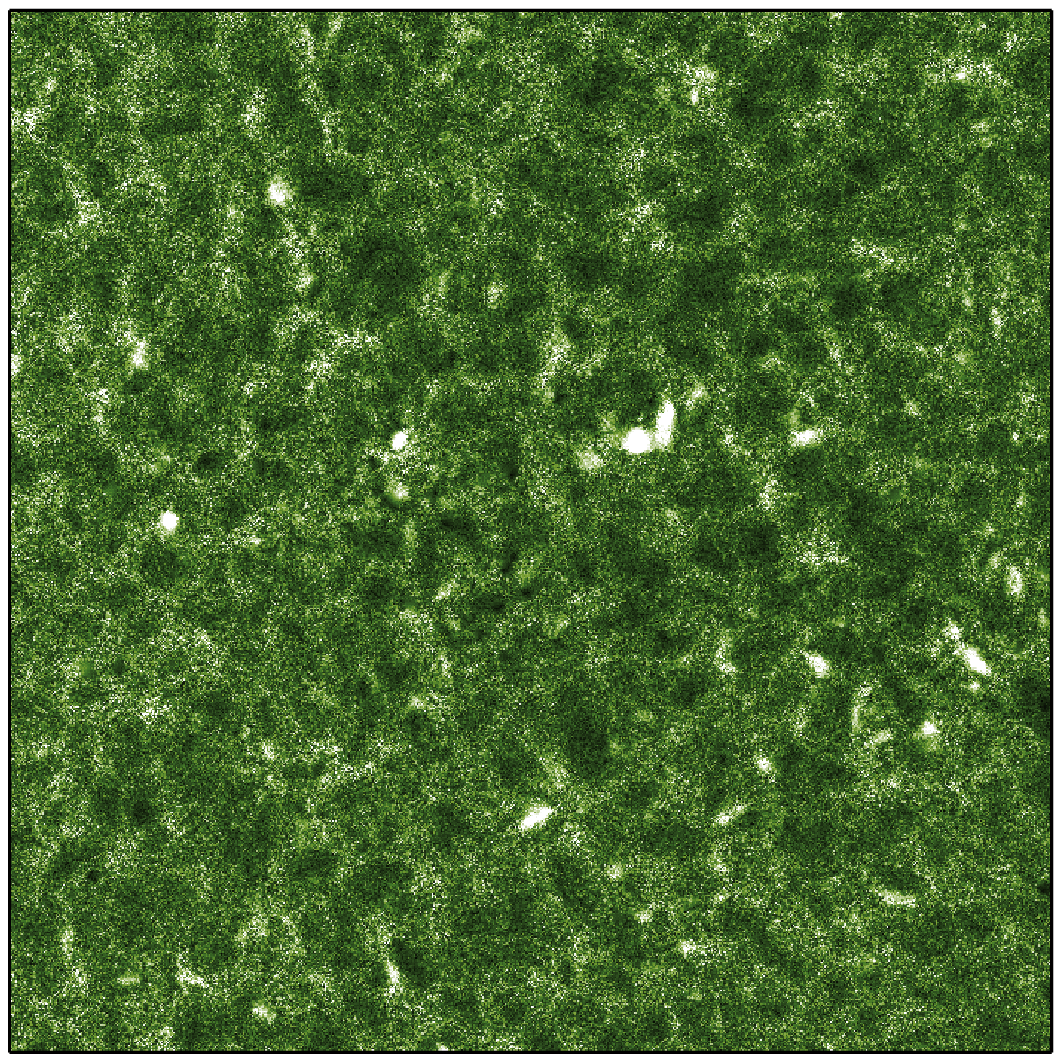}
\hspace{-0.9cm}
\includegraphics[width=0.33\textwidth, bb= 60 40 406 341]{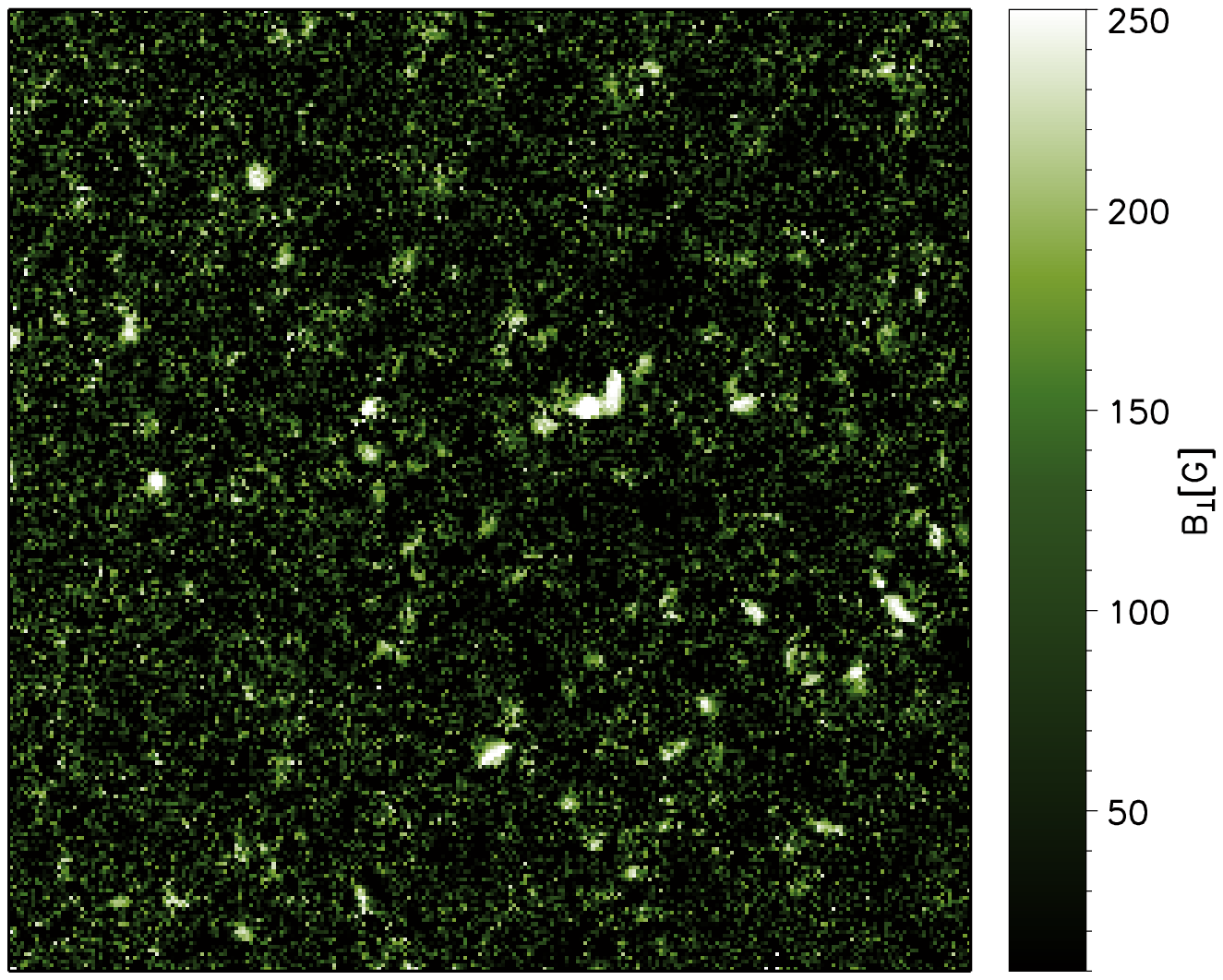}
\caption{Left (centre): inferred value of the longitudinal (transverse) magnetic
field for the  5250.2 \AA\ line. Right: transverse field corrected from the bias
due to pure noise. Central and right panels share the same colorbar.}
\label{fig:imax2}
\end{figure*}

The second experiment consists of filter-polarimetric data observed with the IMaX
instrument on the Sunrise mission. This
polarimeter observes the 5250.2 Fe\,{\sc i} line (for which we have carried out the
bias experiment in Section 4). The data set consists of the four Stokes
parameters observed at the quietest areas of the solar disc centre at a spatial cutoff
frequency of about 0.15-0.18$''$ ($\sim 120$ km in the solar surface, the best
one at the moment for instruments with polarimetric capabilities). The noise
level in circular and linear polarization is $10^{-3}$ in units of the continuum
intensity, I$_\mathrm{c}$. The left panel of Fig. \ref{fig:imax1} shows the
continuum intensity image, with brighter areas associated to granular regions, where
the plasma ascends to the photosphere. Dark areas are the intergranular
lanes, where the motions are preferentially downflowing. The right panel of this same figure displays the estimated
bias for the transversal component of the magnetic field using Eq. \ref{eq:bias}
for the percentile $c=0.5$. The bias has a very particular spatial distribution
which mimics reversed granulation (bright areas become dark and viceversa). 
This particular spectral line is strongly sensitive to the temperature, becoming
very deep and narrow in intergranular lanes and less deep and broad in granules.
The dependence of Eq. \ref{eq:bias} on the second derivative of the intensity profile,
which gives an idea of the width of the spectral line, is the reason why
the bias is larger in intergranules and less important in granules. 

Left panel of Fig. \ref{fig:imax2} displays the inferred longitudinal component
of the magnetic field. As can be seen, the noisy background has values around
zero, consistent with the fact that the estimator of this quantity is unbiased.
The central panel shows the inferred transversal component of the magnetic
field. The noisy background is now filled by magnetic fields that have quite
intense values, illustrating the network pattern of the bias. If this bias for
the longitudinal field is not appropriately account for, this might lead to
an artificial excess of inclined magnetic fields. This effect could be
affecting some of the recent magnetic field inferences in the quiet Sun
observed with Hinode \citep[see, e.g.,][]{orozco_hinode07,lites08,sheminova09,ishikawa09,ishikawa10,ishikawa11}.

Since we can characterize the bias by its median value, it is possible to distinguish
real signals from false signals produced by the presence of noise. The way 
we proceed is as follows. We compute a conservative upper limit for a
trustful field as the bias for the percentile 84 ($c=0.84$ in Eq.
\ref{eq:bias}). This value changes from pixel to pixel. Then, we force $B_\perp=0$ in
those places where the inferred $B_\perp$ is smaller than the bias. The result
is represented in the right panel of Fig. \ref{fig:imax2}. Now, the real signals
(coming from pixels with linear polarization clearly above the noise level) are much 
more evident and most of the background has disappeared. Note that we
have only removed those signals that were produced by the presence of noise (the expected value
of $B_\perp=0$ should be zero). However, if $B_\perp\ne0$, the bias can still be important if
the noise level is high, as shown in Fig. \ref{fig:bias_sigma_cte}.

\section{Conclusions}
We have shown that the weak field approximation (the Stokes parameters are
proportional to the first and second order derivative of the intensity) also holds
for observed stellar fluxes in the case of slow rotators. We have used a
maximum likelihood estimator to infer the magnetic field vector from the
observed Stokes profiles. The main result of this papers is that we give 
explicit formulae for the components of the magnetic field 
vector in terms of the observables. The formulae are general and hold for specific intensities,
for integrated fluxes and are slightly modified for LSD profiles. In the
particular case of a stellar dipole, the orientation of the dipole axis and its
intensity can be recovered from one observation of the full Stokes vector. 

We have also studied the bias of this maximum likelihood estimator. The
longitudinal magnetic field and the azimuthal angle are unbiased quantities.
However, the transversal component of the magnetic field and hence the
inclination of the field are overestimated in the presence of noise. We derive
the estimated value of the perpendicular component of the magnetic field in the
case that there is no linear polarization signal above the noise (the bias for $B_\perp$=0). We
propose to evaluate this bias prior to the inference of the perpendicular component of
the field. One should be very cautious when the inferred $B_\perp$ is of the order of the
bias.

\section*{Acknowledgments}
We are grateful to F. Leone for helpful comments.
This work has been funded by the Spanish Ministry of Science and Innovation
under projects AYA2010-18029 (Solar Magnetism and Astrophysical
Spectropolarimetry) and Consolider-Ingenio 2010 CSD2009-00038.

% \bibliographystyle{mn2e}
% \bibliography{/home/aasensio/Dropbox/biblio.bib}

\bsp

\appendix

\onecolumn
\section{Derivation of the maximum likelihood solution of the stellar magnetic
field}

We start from the general weak-field equations that hold both for the resolved
(solar) case and for the integrated stellar dipole (Eq. 11):
\begin{align}
{\cal V}&=-C {\cal B}_\parallel  {\cal I'} \nonumber \\
{\cal Q}&=-C^2 {\cal B}^2_\perp \cos{2\phi} {\cal I''} \nonumber \\
{\cal U}&=-C^2 {\cal B}^2_\perp \sin{2\phi} {\cal I''},
\end{align}
where the explicit expressions for ${\cal V}, {\cal Q}, {\cal U}, {\cal I'},$
and ${\cal I''},$ for each case can be found in Table \ref{tab:variables} in the
main text. We denote the Stokes vector by $\mathbf{\cal S}=[{\cal I}, {\cal Q},
{\cal U}, {\cal V}]$. Assuming that the difference between the data and the
model follows a Gaussian distribution, the likelihood (the probability
distribution of the data given the parameters) can we written as:
\begin{align}
{\cal L}=\prod_{i=1}^{n_l}\prod_{j=1}^{n_\lambda}\prod_{k=2}^4 \left[2\pi
(\sigma^i_{jk})^2 \right]^{-1/2}\exp\left[
-\frac{({\cal S}^i_{jk}-{\cal S}^{i;\mathrm{mod}}_{jk})^2}{2(\sigma^i_{jk})^2} \right],
\end{align}
where the index $i$ refers to the spectral line, $j$ refers to the wavelength,
and $k$ to the element of the Stokes vector. The abbreviation ``$\mathrm{mod}$'' stands for
the model. 
The symbol $\sigma^2$ stands for 
the variance of ${\cal S}-{\cal S}^{\mathrm{mod}}$. Taking the logarithm, we build the
log-likelihood, $\ln{\cal L}$:
\begin{align}
\ln{\cal L}=-\frac{3n_l n_\lambda}{2}
\ln(2\pi)-\sum_{i=1}^{n_l}\sum_{j=1}^{n_\lambda}\sum_{k=2}^4\left[\frac{({\cal S}^i_{
jk}-{\cal S}^{i;\mathrm{mod}}_{jk})^2}{2(\sigma^i_{jk})^2}+\frac{1}{2}
\ln(\sigma^i_{jk})^2\right].
\end{align}
In order to estimate the parameters that fit the data given the proposed model
we have to maximise the likelihood or, equivalently, minimise $\ln{\cal
L}$. Let $\mathbf{p}=\left[ {\cal B}_\parallel, {\cal B}_\perp, \phi \right]$
denote the set of parameters of our model. Following the standard
approach, to estimate the vector of parameters $\mathbf{p}$, we must solve the following set of equations:
\begin{align}
\frac{\partial \ln{\cal L}}{\partial
p_l}=0=\sum_{i=1}^{n_l}\sum_{j=1}^{n_\lambda}\sum_{k=2}^4\frac{\partial}{
\partial p_l}\left[
\frac{({\cal S}^i_{jk}-{\cal S}^{i;\mathrm{mod}}_{jk})^2}{(\sigma^i_{jk})^2}+\ln(\sigma^i_{jk}
)^2 \right], \qquad \forall l.
\end{align}

There is an important point to clarify given that, in our case, the model is not fully analytical
because it depends on the observed intensity profile \citep{asensio_manso11}. Therefore,
assuming no correlation between the model and the observable (the only difference being
produced by uncorrelated Gaussian noise), the variance for each Stokes parameter is given by:
\begin{align}
(\sigma_2)^2&=\sigma^2({\cal S}_2) + \sigma^2({\cal S}_2^{\mathrm{mod}})=\sigma^2({\cal S}_2) + C^4{\cal
B}^4_\perp \cos^2{2\phi} \, \sigma^2({\cal I''}) \\
(\sigma_3)^2&=\sigma^2({\cal S}_3) + \sigma^2({\cal S}_3^{\mathrm{mod}})=\sigma^2({\cal S}_3) + C^4{\cal
B}^4_\perp \sin^2{2\phi} \, \sigma^2({\cal I''}) \\
(\sigma_4)^2&=\sigma^2({\cal S}_4) + \sigma^2({\cal S}_4^{\mathrm{mod}})=\sigma^2({\cal S}_4) + C^2{\cal
B}^2_\parallel\sigma^2({\cal I'}).
\end{align}
Thanks to the previous equations, the variances that appear in Eq. (A4) depend on the actual parameters, which 
makes the minimisation much harder. Luckily, in the weak field regime and for the observational spectral resolution 
of interest nowadays, it is easy to verify that $\sigma^2({\cal S}_i^{\mathrm{mod}}) \le
\sigma^2({\cal S}_i)$. In any case, this condition should be checked before carrying out
any inversion using the formulae derived in this work. Assuming a first order approximation to the derivative $\mathcal{I'}$,
we find:
\begin{equation}
\sigma^2(\mathcal{I'}) = 2 K^2 \frac{\sigma^2(\mathcal{S}_1)}{\Delta x^2},
\end{equation}
where $K=1$ for the resolved case and $K =\frac{1}{10}\frac{15+u}{6-2u-3v}$ for the dipole case. Assuming
$\sigma^2({\cal S}_i^{\mathrm{mod}}) \le
\sigma^2({\cal S}_i)$ and that the noise in intensity is the same as in the circular
polarisation, it holds that:
\begin{align}
\Delta x \ge \sqrt{2} C {\cal B}_\parallel K \Lambda g,
\end{align}
which means that the spectral sampling has to be larger than or of the order of the
Zeeman splitting. For instance, for a wavelength of $5000$ \AA, a Land\'e
factor of 1.5, and $B_{||}=500$ G, the spectral sampling has to be
larger or equal to 12 m\AA\ for $K=1$. Luckily, this is the case in most of the
observational cases. For example, for the same wavelength, the expected sampling
for two spectrographs with a resolving power of $R=60000, 300000$ are 83 m\AA\
and 16 m\AA, respectively. For the linear polarisation we have, if we assume that their
associated noise equals the noise in intensity, that:
\begin{align}
\Delta x \ge \sqrt{2 G K' \cos{2\phi}} \, C {\cal B}_\perp \Lambda
\nonumber \\
\Delta x \ge \sqrt{2 G K' \sin{2\phi}} \, C {\cal B}_\perp \Lambda,
\end{align}
where $K'=1/4$ for the resolved case and $K' =\frac{1}{4480}\frac{420-68u-105v}{6-2u-3v}$ for the dipole case.
For $B_\perp=500$ G, $\phi=0^\circ$ and $G=2.25$, we end up with $\Delta x=6$ m\AA.

Taking the previous considerations into account, Eq. (A4) can be simplified to:
\begin{align}
\frac{\partial \ln{\cal L}}{\partial
p_l}=0=\sum_{i=1}^{n_l}\sum_{j=1}^{n_\lambda}\sum_{k=2}^4\frac{\partial}{
\partial p_l}\left[
\frac{({\cal S}^i_{jk}-{\cal S}^{i;\mathrm{mod}}_{jk})^2}{(\sigma^i_{jk})^2}\right]\equiv\frac{
\partial \chi^2}{\partial p_l},
\end{align}
where the well known $\chi^2$ merit function is defined as:
\begin{align}
\chi^2=\sum_{ij} \frac{({\cal V}^i_j-{\cal V}^{i;\mathrm{mod}}_j)^2}{(\sigma^i_{{\cal
V}j})^2} &+ \sum_{ij} \frac{({\cal Q}^i_j-{\cal
Q}^{i;\mathrm{mod}}_j)^2}{(\sigma^i_{{\cal Q}j})^2} + \sum_{ij} \frac{({\cal U}^i_j-{\cal
U}^{i;\mathrm{mod}}_j)^2}{(\sigma^i_{{\cal U}j})^2}.
\end{align}
Explicitly, the derivatives of the $\chi^2$ with respect to the parameters we want
to infer are:
\begin{align}
\frac{\partial \chi^2}{\partial {\cal B}_\parallel}&=2C\sum_{ij}\frac{{\cal
V}^i_j
{\cal I'}^i_j}{(\sigma^i_{{\cal V}j})^2} +2C^2{\cal
B}_\parallel\sum_{ij}\frac{({\cal
I'}^i_j)^2}{(\sigma^i_{{\cal V}j})^2} \\
\frac{\partial \chi^2}{\partial {\cal B}_\perp}&=4 C^2 {\cal B}_\perp \left(
\cos{2\phi}\sum_{ij}\frac{{\cal Q}^i_j {\cal I''}^i_j}{(\sigma^i_{{\cal Q}j})^2}
+ \sin{2\phi} \sum_{ij}\frac{{\cal U}^i_j {\cal I''}^i_j}{(\sigma^i_{{\cal
U}j})^2} \right) + 4C^4\mathcal{B}^3_\perp \left( \cos^2{2\phi}\sum_{ij}\frac{{(\cal
I''}^i_j)^2}{(\sigma^i_{{\cal Q}j})^2} + \sin^2{2\phi}\sum_{ij}\frac{({\cal
I''}^i_j)^2}{(\sigma^i_{{\cal U}j})^2} \right) \\
\frac{\partial \chi^2}{\partial \phi}&=4 C^2 \mathcal{B}^2_\perp \left(
\cos{2\phi}\sum_{ij}\frac{{\cal U}^i_j {\cal I''}^i_j}{(\sigma^i_{{\cal U}j})^2}
- \sin{2\phi} \sum_{ij}\frac{{\cal Q}^i_j {\cal I''}^i_j}{(\sigma^i_{{\cal
Q}j})^2} \right) + 4C^4\mathcal{B}^4_\perp \sin{2\phi} \cos{2\phi} \left(
\sum_{ij}\frac{{(\cal I''}^i_j)^2}{(\sigma^i_{{\cal U}j})^2} -
\sum_{ij}\frac{({\cal I''}^i_j)^2}{(\sigma^i_{{\cal Q}j})^2} \right) 
\end{align}
By forcing these derivatives to zero, we obtain the maximum-likelihood estimate of each
parameter. For the longitudinal magnetic field, using Eq. A13, we obtain a
unique solution:
\begin{equation}
{\cal B}_\parallel=-\frac{1}{C}\frac{\sum_{ij} \frac{{\cal V}^i_j {\cal
I'}^i_j}{(\sigma^i_{{\cal
V}j})^2}}{\sum_{ij} \frac{({\cal I'}^i_j)^2}{(\sigma^i_{{\cal V}j})^2}}.
\end{equation}
Dividing Eq. A14 and A15, we obtain a solution for the azimuth:
\begin{equation}
\tan{2\phi}=\frac{\sum_{ij} \frac{({\cal I''}^i_j)^2}{(\sigma^i_{{\cal Q}j})^2}
\sum_{ij} \frac{{\cal U}^i_j{\cal I''}^i_j}{(\sigma^i_{{\cal U}j})^2}
}{\sum_{ij} \frac{({\cal I''}^i_j)^2}{(\sigma^i_{{\cal U}j})^2} \sum_{ij}
\frac{{\cal Q}^i_j{\cal I''}^i_j}{(\sigma^i_{{\cal Q}j})^2} }.
\end{equation}
Dividing Eq. A15 by $\cos{2\phi}$ and using Eq. A17, after some algebra we
obtain two solutions. One is ${\cal B}_\perp=0$, which is not valid since this solution
maximises the $\chi^2$. We have tested this computing the second derivative.
The solution that minimises the $\chi^2$ is given by:
\begin{equation}
{\cal B}_\perp^2=\frac{1}{C^2}\frac{\sqrt{ \left(\sum_{ij}\frac{({\cal
I''}^i_j)^2}{(\sigma^i_{{\cal U}j})^2} \sum_{ij}\frac{{\cal Q}^i_j{\cal
I''}^i_j}{(\sigma^i_{{\cal Q}j})^2}\right)^2 + \left( \sum_{ij}\frac{({\cal
I''}^i_j)^2}{(\sigma^i_{{\cal Q}j})^2} \sum_{ij}\frac{{\cal U}^i_j{\cal
I''}^i_j}{(\sigma^i_{{\cal U}j})^2}\right)^2 }}{\sum_{ij}\frac{({\cal
I''}^i_j)^2}{(\sigma^i_{{\cal U}j})^2}\sum_{ij}\frac{({\cal
I''}^i_j)^2}{(\sigma^i_{{\cal Q}j})^2}}.
\end{equation}

The errors associated with each parameter are computed assuming that the surface around the maximum likelihood
is approximately a multidimensional Gaussian. This is equivalent to assuming a parabolic approximation
to the $\chi^2$ close to the minimum. The curvature close to the minimum is given by the Hessian matrix
$\mbox{\boldmath$\zeta$}$, whose elements are:
\begin{align}
\zeta_{kl}=\frac{1}{2}\frac{\partial^2 \chi^2}{\partial p_k \partial p_l}.
\end{align}
The matrix can be calculated using the following second derivatives:
\begin{align}
\frac{\partial^2 \chi^2}{\partial {\cal
B}_\parallel^2}&=2C^2\sum_{ij}\frac{({\cal
I'}^i_j)^2}{(\sigma^i_{{\cal V}j})^2} \\
\frac{\partial^2 \chi^2}{\partial {\cal B}_\perp^2}&=8C^4{\cal B}_\perp^2\left(
\cos^2{2\phi}\sum_{ij}\frac{({\cal I''}^i_j)^2}{(\sigma^i_{{\cal Q}j})^2} +
\sin^2{2\phi}\sum_{ij}\frac{({\cal I''}^i_j)^2}{(\sigma^i_{{\cal U}j})^2}
\right) \\
\frac{\partial^2 \chi^2}{\partial \phi^2}&=8C^4{\cal B}_\perp^4\left(
\sin^2{2\phi}\sum_{ij}\frac{({\cal I''}^i_j)^2}{(\sigma^i_{{\cal Q}j})^2} +
\cos^2{2\phi}\sum_{ij}\frac{({\cal I''}^i_j)^2}{(\sigma^i_{{\cal U}j})^2}
\right)\\
\frac{\partial \chi^2}{\partial {\cal B}_\parallel \partial {\cal
B}_\perp}&=\frac{\partial
\chi^2}{\partial {\cal B}_\perp \partial {\cal B}_\parallel}=0\\
\frac{\partial \chi^2}{\partial {\cal B}_\parallel \partial\phi}&=\frac{\partial
\chi^2}{\partial\phi \partial {\cal B}_\parallel}=0\\
\frac{\partial \chi^2}{\partial {\cal B}_\perp \partial \phi}&=\frac{\partial
\chi^2}{\partial \phi\partial {\cal B}_\perp}=8C^4{\cal
B}_\perp^3\sin{2\phi}\cos{2\phi}\left(
\sum_{ij}\frac{({\cal I''}^i_j)^2}{(\sigma^i_{{\cal U}j})^2}
-\sum_{ij}\frac{({\cal I''}^i_j)^2}{(\sigma^i_{{\cal Q}j})^2} \right).
\end{align}

The square root of the diagonal of the covariance matrix gives the error estimates for each
parameter. Such covariance matrix is just the inverse of the Hessian matrix, 
$\mbox{\boldmath$\mathcal{C}$}=\mbox{\boldmath$\zeta$}^{-1}$. In our case:
\begin{equation}
\mbox{\boldmath$\mathcal{C}$} = \left[ 
\begin{array}{ccc}
\mathcal{C}_{11} & 0 & 0 \\
0 & \mathcal{C}_{22} & \mathcal{C}_{23} \\
0 & \mathcal{C}_{32} & \mathcal{C}_{33}\end{array} \right],
\end{equation}
with 
\begin{align}
\mathcal{C}_{11}&=\left[ {C^2 {\sum_{ij} \frac{({\cal
I'}^i_j)^2}{(\sigma^i_{{\cal V}j})^2}}} \right]^{-1}\\
\mathcal{C}_{22}&=\frac{1}{4C^4{\cal B}_\perp^2} { \frac{\sin^2{2\phi} \sum_{ij}
\frac{({\cal I''}^i_j)^2}{(\sigma^i_{{\cal Q}j})^2} +\cos^2{2\phi} \sum_{ij}
\frac{({\cal I''}^i_j)^2}{(\sigma^i_{{\cal
U}j})^2}}{\cos^2{2\phi}\sin^2{2\phi}\left( \sum_{ij} \frac{({\cal
I''}^i_j)^2}{(\sigma^i_{{\cal Q}j})^2} -\sum_{ij} \frac{({\cal
I''}^i_j)^2}{(\sigma^i_{{\cal U}j})^2}  \right)^2 +\sum_{ij} \frac{({\cal
I''}^i_j)^2}{(\sigma^i_{{\cal Q}j})^2} \sum_{ij} \frac{({\cal
I''}^i_j)^2}{(\sigma^i_{{\cal U}j})^2}}}\\
\mathcal{C}_{23}&=\mathcal{C}_{32}=\frac{1}{4C^4{\cal
B}_\perp^{3}}{\frac{\sin{2\phi}\cos{
2\phi} \left( \sum_{ij} \frac{({\cal I''}^i_j)^2}{(\sigma^i_{{\cal U}j})^2}
-\sum_{ij} \frac{({\cal I''}^i_j)^2}{(\sigma^i_{{\cal Q}j})^2} 
\right)^2}{\cos^2{2\phi}\sin^2{2\phi}\left( \sum_{ij} \frac{({\cal
I''}^i_j)^2}{(\sigma^i_{{\cal Q}j})^2} -\sum_{ij} \frac{({\cal
I''}^i_j)^2}{(\sigma^i_{{\cal U}j})^2}  \right)^2 +\sum_{ij} \frac{({\cal
I''}^i_j)^2}{(\sigma^i_{{\cal Q}j})^2} \sum_{ij} \frac{({\cal
I''}^i_j)^2}{(\sigma^i_{{\cal U}j})^2}}}\\
\mathcal{C}_{33}&=\frac{1}{4C^4{\cal B}_\perp^4}{\frac{\cos^2{2\phi} \sum_{ij}
\frac{({\cal I''}^i_j)^2}{(\sigma^i_{{\cal Q}j})^2} +\sin^2{2\phi} \sum_{ij}
\frac{({\cal I''}^i_j)^2}{(\sigma^i_{{\cal
U}j})^2}}{\cos^2{2\phi}\sin^2{2\phi}\left( \sum_{ij} \frac{({\cal
I''}^i_j)^2}{(\sigma^i_{{\cal Q}j})^2} -\sum_{ij} \frac{({\cal
I''}^i_j)^2}{(\sigma^i_{{\cal U}j})^2}  \right)^2 +\sum_{ij} \frac{({\cal
I''}^i_j)^2}{(\sigma^i_{{\cal Q}j})^2} \sum_{ij} \frac{({\cal
I''}^i_j)^2}{(\sigma^i_{{\cal U}j})^2}}}
\end{align}

The error bars are thus given by:
\begin{equation}
\delta p_k = \pm \Delta \sqrt{\mathcal{C}_{kk}},
\end{equation}
with $\Delta=1, 1.65, 2, 2.57, 3,$ and $3.89$ for a confidence level of 68.3\%,
90\%, 95.4\%, 99\%, 99.73\%, and 99.99\%, respectively. This computation assumes
that, to estimate the error of a parameter, we fix the values of the rest to 
the ones that maximize the likelihood and compute the confidence levels for
the one-dimensional probability distribution of this parameter \citep[see][for
more details]{numerical_recipes86}. Note that this approximation does not take
into account the degeneracies between parameters since it does not integrate the
probability distribution of the rest of the parameters but fixes a certain value
\citep[see][for a robust Bayesian inversion]{asensio_bayeswf11}.
Note also that the covariance matrix is diagonal if the standard deviation of ${\cal Q}$ and
${\cal U}$ are the same ($\sigma^i_{{\cal Q}j}=\sigma^i_{{\cal U}j}$). This is
generally the case in solar observations in which the measurement efficiencies for linear polarization
are similar.

\label{lastpage}

\end{document}